\documentclass[floats,floatfix,showpacs,amssymb,prd,twocolumn,superscriptaddress,nofootinbib]{revtex4-1}

\usepackage{graphicx,epsf, epsfig, amssymb}
\usepackage{bm}
\usepackage{longtable}
\usepackage[usenames,dvipsnames]{xcolor}
\usepackage{color}
\usepackage[colorlinks=true,linktocpage=false,citecolor=blue,linkcolor=blue,urlcolor=blue]{hyperref}
\usepackage{amsfonts,amsmath,amssymb,mathrsfs}
\usepackage{multirow}
\usepackage[nolist,nohyperlinks]{acronym}
\usepackage{xspace}
\usepackage{booktabs,array}

\def\etal{et al.\xspace}
\def\bns{\ac{NS}-\ac{NS}\xspace}
\def\bbh{\ac{BH}-\ac{BH}\xspace}
\def\nsbh{\ac{NS}-\ac{BH}\xspace}
\def\sacra{{\ttfamily SACRA}\xspace}

\def\be{\begin{equation}}
\def\ee{\end{equation}}
\def\beq{\begin{eqnarray}}
\def\eeq{\end{eqnarray}}
\def\ben{\begin{enumerate}}
\def\een{\end{enumerate}}
\def\bi{\begin{itemize}}
\def\ei{\end{itemize}}
\def\f{\frac}

\newcommand{\nn}{\nonumber}

\newcommand{\fCut}{f_{\rm cut}}
\newcommand{\spin}{\chi}
\newcommand{\mBH}{M_{\rm BH}}
\newcommand{\mNS}{M_{\rm NS}}
\newcommand{\mSun}{M_\odot}
\newcommand{\bmNS}{\bar{M}_{\rm NS}}
\newcommand{\rNS}{R_{\rm NS}}
\newcommand{\spinf}{\chi_{\rm f}}
\newcommand{\mBHf}{M_{\rm f}}
\newcommand{\fRD}{f_{\rm RD}}
\newcommand{\tfRD}{\tilde{f}_{\rm RD}}
\newcommand{\fTide}{f_{\rm tide}}
\newcommand{\mTorus}{M_{\rm b, torus}}

\begin{document}

\title{Aligned spin neutron star-black hole mergers: \\ a
  gravitational waveform amplitude model}

\author{Francesco Pannarale}
\email{francesco.pannarale@ligo.org}
\affiliation{School of Physics and Astronomy, Cardiff University, The
  Parade, Cardiff CF24 3AA, UK}

\author{Emanuele Berti}
\email{eberti@olemiss.edu}
\affiliation{Department of Physics and Astronomy, The University of
  Mississippi, University, MS 38677, USA}
\affiliation{CENTRA, Departamento de F\'isica, Instituto Superior
  T\'ecnico, Universidade de Lisboa, Avenida Rovisco Pais 1, 1049
  Lisboa, Portugal}

\author{Koutarou Kyutoku} 
\email{koutarou.kyutoku@riken.jp}
\affiliation{Interdisciplinary Theoretical Science (iTHES) Research
  Group, RIKEN, Wako, Saitama 351-0198, Japan}

\author{Benjamin D. Lackey} 
\email{bdlackey@syr.edu}
\affiliation{Department of Physics, Princeton University, Princeton,
  NJ 08544, USA}
\affiliation{Department of Physics, Syracuse University, Syracuse, NY
  13244, USA}

\author{Masaru Shibata}
\email{mshibata@yukawa.kyoto-u.ac.jp}
\affiliation{Yukawa Institute for Theoretical Physics, Kyoto
  University, Kyoto 606-8502, Japan}


\date{\today}

\begin{abstract}
  The gravitational radiation emitted during the merger of a black
  hole with a neutron star is rather similar to the radiation from the
  merger of two black holes when the neutron star is not tidally
  disrupted. When tidal disruption occurs, gravitational waveforms can
  be broadly classified in two groups, depending on the spatial extent
  of the disrupted material. Extending previous work by some of us,
  here we present a phenomenological model for the gravitational
  waveform amplitude in the frequency domain encompassing the three
  possible outcomes of the merger: no tidal disruption, ``mild'' and
  ``strong'' tidal disruption. The model is calibrated to 134
  general-relativistic numerical simulations of binaries where the
  black hole spin is either aligned or antialigned with the orbital
  angular momentum. All simulations were produced using the \sacra
  code and piecewise polytropic neutron star equations of state. The
  present model can be used to determine when black-hole binary
  waveforms are sufficient for gravitational-wave detection, to
  extract information on the equation of state from future
  gravitational-wave observations, to obtain more accurate estimates
  of black hole-neutron star merger event rates, and to determine the
  conditions under which these systems are plausible candidates as
  central engines of gamma-ray bursts, macronov\ae~and kilonov\ae.
\end{abstract}

\maketitle

\begin{acronym}
\acrodef{BH}[BH]{black hole}
\acrodef{EM}[EM]{electromagnetic}
\acrodef{EOB}[EOB]{effective-one-body}
\acrodef{EOS}[EOS]{equation of state}
\acrodef{GW}[GW]{gravitational-wave}
\acrodef{IMR}[IMR]{inspiral-merger-ringdown}
\acrodef{ISCO}[ISCO]{innermost stable circular orbit}
\acrodef{KAGRA}[KAGRA]{Kamioka Gravitational wave detector}
\acrodef{LIGO}[LIGO]{Laser Interferometer Gravitational-Wave Observatory}
\acrodef{NS}[NS]{neutron star}
\acrodef{PN}[PN]{post-Newtonian}
\acrodef{QNM}[QNM]{quasinormal mode}
\acrodef{SGRB}[SGRB]{short gamma-ray burst}
\end{acronym}

\section{Introduction}
The year 2015 will mark the beginning of the advanced \ac{GW} detector
era.  Exactly one hundred years after Einstein formulated the theory
of General Relativity, the two Advanced \ac{LIGO} detectors
\cite{TheLIGOScientific:2014jea, 2010CQGra..27h4006H} are about to
start their observation runs.  They will soon be followed by Virgo
\cite{TheVirgo:2014hva}, and later on by the \ac{KAGRA}
\cite{Somiya:2011np, AsoKAGRA} and \ac{LIGO}-India \cite{indigo}.
Detections will provide us with unprecedented information about
astrophysical \ac{GW} sources. Coalescing compact binary systems
containing \acp{NS} and/or \acp{BH} are the main target for \ac{GW}
interferometric detectors.  Their waveforms encode information about
the masses, spins, distance, sky location, and orientation of the
source, and, when \acp{NS} are present, about the \ac{NS} \ac{EOS}.
Detecting \acp{GW} emitted by compact binaries relies on matching
noisy detector data with theoretical signal predictions and,
therefore, requires us to build waveforms for the targeted sources
that are as accurate as possible.  At the same time, interpreting
future observations calls for understanding as many details as
possible about the sources.

In light of this, numerical relativity has made giant steps forward
over the last decade, and simulations of the late inspiral and merger
of compact binaries are now possible.  As these calculations are
resource intensive and time consuming, simulations that cover as many
cycles as are necessary to fill the sensitivity bandwidth of the
detectors and that span the whole parameter space are still beyond the
reach of present-day computers.  This is why semianalytical waveform
models that fill the gap between perturbative methods --- that
describe the early inspiral stage --- and numerical relativity are
necessary.  These models and numerical simulations are most advanced
for \bbh systems. Phenomenological \ac{IMR} waveform models have been
proposed for nonspinning binaries by Ajith \etal \cite{Ajith:2007qp,
  Ajith:2007kx, Ajith:2007xh}; for spinning, nonprecessing binaries by
Ajith \etal \cite{Ajith:2009bn} and by Santamar{\'{\i}}a, Ohme, \etal
\cite{Santamaria:2010yb}, and more recently by Khan \etal
\cite{PhenomDa, PhenomDb}; and for spinning, precessing binaries by
Hannam \etal \cite{PhenomP}.  These are generally referred to as
``PhenomA,'' ``PhenomB,'' ``PhenomC,'' ``PhenomD,'' and ``PhenomP,''
respectively, and are all based on a \ac{PN} description of the early
inspiral.  Similarly, a lot of effort was put into tuning the \ac{EOB}
model to \bbh numerical simulations, and progress in this direction
has been remarkable (see e.g.~\cite{Taracchini:2012ig, Damour:2012ky,
  Damour:2010zb, Ohme:2011zm, Pan2014, Taracchini2014, Szilagyi2015}).
\ac{IMR} models are employed in several contexts: to design and build
\ac{GW} detection templates at manageable computational costs
(e.g.~\cite{Brown2013,Harry2014}), to test \ac{GW} detection
infrastructure (e.g.~\cite{Aylott:2009ya, Ajith:2012az, NRAR2013}), to
evaluate statistical and systematic errors (e.g.~\cite{Ajith:2009fz,
  Ohme2013}) and to perform detection rate calculations
(e.g.~\cite{Abadie:2010cf, Marassi:2011si, Zhu:2012xw, Dominik2014}).

In the case of compact binaries containing at least one \ac{NS}, the
waveform modelling landscape is less developed, both because long and
accurate simulations are particularly hard to achieve and because the
parameter space is larger. The outcome and aftermath of \bns and \nsbh
binary simulations (as opposed to \bbh binaries) depends on several
assumptions on physics that is currently underconstrained, including
e.g.~the \ac{NS} \ac{EOS}, the effect of magnetic fields and neutrino
emission.  This makes the \bns case particularly complex, as a
hypermassive \ac{NS} may form in the merger and oscillate for
$10$--$100$ milliseconds, emitting \acp{GW} in a fashion that is hard
to predict from the parameters of the binary
itself~\cite{SekiguchiPRL2011, Hotokezaka:2011dh, Hotokezaka2013}.
Due to this complex late stage of the evolution, attempts at
constructing \ac{EOB}-based waveform models valid up to merger
\cite{Bernuzzi:2014owa} and possibly beyond
\cite{Bauswein:2015yca,Bernuzzi:2015rla} are still in their infancy.
For \nsbh systems --- the focus of this paper --- one expects
relatively large mass ratios, which cause complications at both the
analytical and numerical level: the convergence of the analytical
\ac{PN} approximation is expected to be slower than for \bns systems
\cite{Buonanno:2009zt}, residual eccentricity in the initial data can
be appreciable \cite{Berti:2007cd,Kyutoku:2009sp,Kyutoku:2014yba} (but
see also \cite{foucart_kpt2008,Henriksson:2014tba}), and very
different dynamical time scales must be tracked by numerical evolution
codes~\cite{Shibata:2009cn, Kyutoku:2010zd, Kyutoku:2011vz,
  Foucart:2012vn, Kyutoku2013, Foucart:2013, Deaton2013, Tanaka2014,
  Foucart2014, Foucart2015, Kyutoku2015, Etienne:2008re, Duez09,
  Etienne2012, Etienne2012b, Lovelace2013, Paschalidis2013b,
  Kawaguchi:2015bwa, Kiuchi2015}.

Despite these obstacles, a clear picture of the \ac{GW} emission of
\nsbh binaries has emerged over the last few years.  Most of the
\ac{GW} signal is emitted before the \ac{NS} is tidally disrupted ---
if this happens at all --- and before significant thermal effects
occur.  Furthermore, magnetic fields appear to barely affect \ac{GW}
emission~\cite{Etienne2012b}.  These are particularly fortunate
circumstances, as they imply that an ideal fluid-dynamics treatment
with a cold \ac{EOS} and an ideal-gas $\Gamma$-law for the thermal
part are appropriate to simulate the dynamical regime that is of
interest for the \ac{GW} signal~\cite{Deaton2013}.  At least two
papers attempted a phenomenological description of the \ac{GW}s
emitted by \nsbh binaries.

In the first paper, Lackey \etal \cite{Lackey2014} developed an
analytic representation of the \nsbh \ac{IMR} waveform calibrated to
$134$ numerical waveforms produced by the \sacra code~\cite{SACRA}
with the main goal of assessing the measurability of the \ac{NS} tidal
deformability.

A subset of these simulations for systems with non-spinning \acp{BH}
was then used in work by Pannarale \etal \cite{PaperI} (henceforth
Paper I) to obtain a phenomenological \nsbh \ac{IMR} waveform
\emph{amplitude} model in the frequency domain.  This model was, at
heart, a ``distortion'' of the PhenomC \bbh model.  Paper I paid
particular attention to the accuracy of the \ac{GW} spectrum at high
frequencies --- where the \ac{EOS}-related phenomenology takes place
--- and to the determination of a cutoff frequency in the \ac{GW}
emission.  This cutoff frequency is especially important in the
construction of \nsbh template banks.  If a \bbh-like template built
to detect a disruptive \nsbh coalescence were to be truncated at a
frequency that is too low with respect to the physical cutoff
frequency of the source, a loss in recovered signal-to-noise ratio
would occur.  If on the other hand the truncation frequency were to be
increased in order to counteract this problem, it could become too
high with respect to the physical cutoff frequency of the \nsbh
source, and this would possibly result in penalizing the template by
degrading its performance in chi-square tests, which would also be
detrimental to the detection.

The goal of the present paper is to extend the work of Paper I to
\nsbh systems with a non-precessing, spinning \ac{BH}, using the full
set of $134$ hybrid waveforms considered in~\cite{Lackey2014}.  The
phenomenological model based on this catalog allows us to produce the
most accurate determination of cutoff frequencies for \nsbh \ac{GW}
signals, with relative errors on the cutoff frequency below $10$\%.
These errors are well below the errors one would obtain using either
\bbh models or the \nsbh model of \cite{Lackey2014}, with immediate
applications in setting up template banks to target these systems.  As
for Paper I, we adopt a conservative approach and focus on the
analytical modeling of the \ac{GW} {\em amplitude} in the frequency
domain, because residual eccentricity in our initial data and the
short duration of our simulations do not guarantee an accurate phasing
in the whole parameter space: see Hannam \etal \cite{Hannam2010} for
how mass ratio affects the minimum number of numerical waveform cycles
necessary to ensure an accurate phase and amplitude modelling, and
\cite{Lackey2014} for issues in building hybrid waveforms for \nsbh
binaries.

The plan of the paper is as follows.  In Secs.~\ref{sec:NRdata} and
\ref{sec:hybrids} we review the basics of the numerical simulations
and gravitational waveform hybrids, respectively, used to build and
test the phenomenological model discussed in this paper.  In
Sec.~\ref{sec:PhenoMixedBGWfs} we describe the waveform model for
\nsbh binaries with a spinning \ac{BH}.  In Sec.~\ref{sec:tests} we
compare our model against numerical data. Sec.~\ref{sec:Applications}
discusses some important applications of our model, in particular
predictions for the tidal disruption frequency and their implications
for \ac{GW} detection and the modeling of \acp{SGRB}.  Finally, in
Sec.~\ref{sec:conclusions} we summarize our conclusions and point out
directions for future work.  Throughout the paper, unless otherwise
noted, we use geometrical units ($G=c=1$).

\section{The Numerical Simulations}\label{sec:NRdata}
Our phenomenological models are calibrated to and tested against the
gravitational waveforms used in \cite{Lackey2014}. The waveforms are
derived by numerical-relativity simulations performed by the \sacra
adaptive-mesh refinement code~\cite{SACRA}. The details of the code
are described in \cite{Kyutoku:2011vz}. Here we only briefly discuss
the key differences with respect to the simulations performed to
derive the waveforms used in Paper I.  Binaries in quasiequilibrium
states are prepared as initial conditions for the simulations using
the multidomain spectral method library {\ttfamily LORENE}
\cite{LORENE}. In this work we allow the \acp{BH} to have nonzero
spins aligned with the orbital angular momentum of the binary.  The
formulation and numerical methods for computing quasiequilibrium
configurations are the same as in \cite{2009PhRvD..79l4018K}, except
for the implementation of \ac{BH} spins \cite{Kyutoku:2011vz}.
Gravitational waveforms are computed from the Weyl scalar $\Psi_4$ by
integrating twice in time using a so-called fixed-frequency
integration method \cite{2011CQGra..28s5015R} to filter out unphysical
low-frequency components (see also \cite{Lackey:2011vz,Lackey2014}).

We adopt piecewise polytropic \acp{EOS}, which mimic
nuclear-theory-based \acp{EOS} with a small number of parameters
\cite{Read:2008iy}, to model the \ac{NS} matter at zero temperature.
Each piecewise polytrope is characterized by polytropic constants
$\kappa_i$ and adiabatic indices $\Gamma_i$ as
\be
P(\rho)=\kappa_i \rho^{\Gamma_i}\quad {\rm for}\quad \rho_{i-1}\leq
\rho <\rho_i \quad (i=1\,\dots\,,n)\,,
\ee
where $\rho$ and $P$ are the rest-mass density and pressure,
respectively. At the critical densities $\rho_i$ we further require
the pressure to be continuous, i.e.
\be
\kappa_i \rho_i^{\Gamma_i}=\kappa_{i+1} \rho_i^{\Gamma_{i+1}}\,,
\ee
and the \ac{EOS} is thus completely specified by $\kappa_1$,
$\Gamma_i$, and $\rho_i$ ($i=1\,\dots\,,n$).  In this work, we adopt
the same piecewise polytropes that were adopted in Paper I and in
\cite{Lackey2014}.  More specifically, $n$ is set to be $2$, the
parameters $\{ \kappa_1 , \Gamma_1 \}$ for the low-density crust
regions are fixed, and the two parameters $\Gamma_2$ and
$P_\mathrm{fidu} \equiv P ( \rho = 10^{14.7}~\mathrm{g/cm}^3 )$ are
systematically varied to span a plausible range of nuclear-matter
properties.  In the dynamical simulations, thermal corrections are
added in an ideal-gas-like form in order to capture the effect of
shock heating \cite{Kyutoku:2010zd,Kyutoku:2011vz}.

The $134$ \nsbh simulations used in this paper are listed in Table II
of \cite{Lackey2014}.  The mass ratio $Q\equiv\mBH/\mNS$ spans the
values $\{2,3,4,5\}$ and the \ac{BH} dimensionless spin parameter
$\chi$ takes the values $\{-0.5,0,0.25,0.5,0.75\}$.  We adopt
$\chi = -0.5$ only for $Q=2$, because the combination of negative
$\chi$ and large $Q$ (say $\ge 3$) yields small tidal effects during
the coalescence.  The \ac{NS} mass $\mNS$ is set to $1.35M_\odot$ for
all the runs, with the exception of some $(Q,\chi) = (2,0.75)$ and
$(2,0)$ cases, in which $\mNS$ can also take the values
$\{1.20M_\odot,1.45M_\odot\}$ and $1.45M_\odot$, respectively.  The
\acp{EOS} are the same $21$ models used in Paper I and
\cite{Lackey2014} (see Fig.\,1 therein for a representation in the
piecewise polytropic \ac{EOS} parameter space) and Fig.\,1
in~\cite{Kyutoku:2010zd} for the \ac{NS} equilibrium sequences yielded
by these \acp{EOS}).  For all combinations of $Q$ and $\chi$ with
$\mNS=1.35M_\odot$, the runs are performed adopting \acp{EOS} with
$\Gamma_2 = 3.0$; additionally, \acp{EOS} with $\Gamma_2 = 2.4$,
$2.7$, and $3.3$ are employed for models with $(Q,\chi) = (2,0)$,
$(3,0.5)$, and $(5,0.75)$.

To build and test our phenomenological \nsbh waveform model we use the
hybrid waveforms of \cite{Lackey2014}, which are also based on the
numerical-relativity simulations just described (see next section).
As in Paper I, we divide the datasets into two groups: one to build
the waveform model and one to test it.  In the simulations used to
build the model the \ac{NS} mass is $\mNS=1.35M_\odot$, and we use the
$\Gamma_2 = 3.0$ \acp{EOS} denoted by 2H, H, HB, and B with
$\log(P_\mathrm{fidu}/(\mathrm{dyne/cm}^{2}))=34.9, 34.5, 34.4$, and
$34.3$, respectively (see, for example, Paper I for this
nomenclature): these are $59$ datasets.  The remaining $75$ cases are
not used to tune the waveform model, but just to test it.

\section{The Hybrid Waveforms}\label{sec:hybrids}
In order to build our phenomenological \nsbh frequency-domain waveform
amplitude model, we must first construct accurate \ac{IMR} waveforms.
This is done by matching each of the numerical \nsbh waveforms
described in the previous section --- which all begin $\sim 10$
\ac{GW} cycles before merger --- to an inspiral waveform model, and by
then splicing them together.  We use the PhenomC \bbh model of
\cite{Santamaria:2010yb} as our inspiral waveform --- a sound
approximation, as tidal effects on the amplitude are negligible in
this stage --- and, unless otherwise noted and in accordance with the
conventions of \cite{Santamaria:2010yb}, all frequencies in this
section and in the rest of the paper are to be intended as multiplied
by the sum $m_0=\mNS+\mBH$ of the two initial masses, i.e.~we use
units in which $m_0=1$.  Similarly, times are to be intended as
divided by $m_0$.

When matching waveforms, a time constant $\tau$ and phase constant
$\phi$ are the two free parameters that need to be fixed.  For a
generic waveform $h(t)$, the time and phase can be adjusted to produce
a shifted waveform $h^{\rm shift}(t; \tau, \phi) = h(t-\tau)e^{i
  \phi}$.  The Fourier-transformed waveform\footnote{We omit the tilde
  over Fourier-transformed quantities in order to keep the notation
  lighter, as $h(t)$ will no longer appear in the rest of the paper.},
which can be written in terms of amplitude and phase as $h(f) = A(f)
e^{i\Phi(f)}$, has a corresponding shifted waveform $h^{\rm shift}(f;
\tau, \phi) = A(f) e^{i\Phi^{\rm shift}(f; \tau, \phi)}$, where
$\Phi^{\rm shift}(f; \tau, \phi) = \Phi(f) + 2\pi f \tau +
\phi$. Because the time and phase constants have no impact on the
\ac{GW} amplitude in the frequency domain, we do not need to calculate
them in this work, where we only model the amplitude $A(f)$.

Although the time and phase constants do not impact the amplitude,
there is still some freedom in constructing the hybrid. We will use
the method of Lackey \etal \cite{Lackey2014}.  In the time domain, the
numerical waveform begins with a finite amplitude, leading to the
oscillatory Gibbs phenomenon that results from Fourier transforming a
waveform segment with nonzero starting amplitude.  We therefore begin
by windowing the numerical \nsbh waveform with a Hann window over the
interval $w_i$ to $w_f$ (of width $w_f - w_i$) defined as
\be
\label{eq:window}
w_{\rm on}(t) = \frac{1}{2}\left[1-\cos\left(\frac{\pi [t - w_i]}{w_f
      - w_i}\right)\right].
\ee
We set the start of the window to be the start of the numerical
waveform at $w_i = 0$ and use a width of $300$ by setting $w_f=300$,
as in Ref.~\cite{Lackey2014}.

We are also free to choose the frequency interval for splicing the
numerical waveform onto the analytic, inspiral waveform.  This
interval should be at high enough frequencies to exclude the effects
of windowing at the beginning of the numerical waveform.  It should
also exclude the small initial eccentricity ($e_0\sim0.03$) that dies
down after a few orbits and results from providing the numerical
simulation with quasicircular (zero radial velocity) initial
conditions that ignore the small radial velocity due to \ac{GW}
radiation reaction.  However, the splicing interval should also be at
a frequency that is low enough to capture the matter effects, present
in the numerical simulations, that take place just before merger. We
smoothly turn on the numerical waveform and smoothly turn off the
analytic, inspiral waveform within a splicing window $s_i < f < s_f$
using Hann windows:
\begin{align}
  \label{eq:splice}
  w_{\rm off}(f) &= \frac{1}{2}\left[1+\cos\left(\frac{\pi [f - s_i]}{s_f - s_i}\right)\right],\\
  w_{\rm on}(f) &= \frac{1}{2}\left[1-\cos\left(\frac{\pi [f -
        s_i]}{s_f - s_i}\right)\right].
\end{align}
The amplitude of the hybrid waveform is then
\begin{align}
  &A_{\rm hybrid}(f) =\nonumber\\
  &\left\{\begin{array}{lc}
      A_{\rm BBH}(f), & \, f \le s_i, \\
      w_{\rm off}(f) A_{\rm BBH}(f) + w_{\rm on}(f) A_{\rm NR}(f), & \, s_i < f \le s_f, \\
      A_{\rm NR}(f), & \, f>s_f.
    \end{array}\right.
\end{align}
As in Ref.~\cite{Lackey2014}, we use a starting frequency of
$s_i=0.018$ and an ending frequency of $s_f=0.019$.

\section{Modeling Spinning Neutron Star-Black Hole
  Waveforms}\label{sec:PhenoMixedBGWfs}
In this section we provide a detailed description of our
phenomenological model for the frequency-domain \ac{GW} amplitude of
nonprecessing \nsbh binaries with a spinning \ac{BH} component.  This
new model generalizes the model presented in Paper I for nonspinning
binaries; throughout the discussion we will point out differences with
respect to the formulation reported in Paper I.

In accordance with the simulations and hybrid waveforms at our
disposal, we set the \ac{BH} spin vector to be aligned to the orbital
angular momentum of the binary.  Additionally, we use the notation
$w_{{f_0,d}}^\pm(f)$ for the windowing functions
\beq
w_{{f_0,d}}^\pm(f)\equiv \f{1}{2}\left[ 1\pm
  \tanh\left(\f{4(f-f_0)}{d}\right)\right]
\eeq
centered in $f_0$ with width $d$.

Before discussing the waveform model itself, we must introduce two
reference \ac{GW} frequencies: these are $\fTide$ and $\fRD$.  The
former is the \ac{GW} frequency at the \emph{onset} of the \ac{NS}
tidal disruption, while the latter is the dominant ($\ell=m=2$, $n=0$)
ringdown frequency of the remnant \ac{BH}. The \ac{BH} remnant
dominant ringdown frequency $\fRD$ depends on the the mass $\mBHf$ and
spin parameter $\spinf$ of the \ac{BH} remnant of the \nsbh merger.
These are calculated according to the model discussed
in~\cite{Pannarale2012, Pannarale2014}, while the fitting formulas
that relate $\spinf$ and $\mBHf$ to $\fRD$ are provided
in~\cite{BertiCardosoWill}.  To calculate the parameters of the
\ac{BH} remnant given the initial parameters of the binary, we follow
the model detailed in~\cite{Pannarale2012, Pannarale2014}.  To compute
$\fTide$, on the other hand, one must first determine a coefficient
$\xi_\text{tide}$ that provides a relativistic correction to the
standard Newtonian estimate of the orbital radius at
\emph{mass-shedding}~\cite{Foucart:2012nc}.  This coefficient can be
found by solving the equation
\be%
\label{eq:xi-tide}%
\f{M_\text{NS}\xi_\text{tide}^3}{M_\text{BH}} =
\f{3[\xi_\text{tide}^2-2\mu\xi_\text{tide}+\mu^2\spin^2]}{\xi_\text{tide}^2-3\mu\xi_\text{tide}+2\spin\sqrt{\mu^3\xi_\text{tide}}}\,,%
\ee%
where $\mu=\mBH/\rNS=Q\mathcal{C}$, with $\mathcal{C}=\mNS/\rNS$
denoting the \ac{NS} compactness.  The orbital radius at mass-shedding
is now given by
\begin{align}%
  \label{eq:rtide}
  \tilde{r}_\text{tide}=\xi_\text{tide}\rNS(1-2\mathcal{C})\,.
\end{align}
The tidal frequency $\fTide$ then reads
\begin{align}
  \label{eq:ftide}
  \fTide=\pm\f{1}{\pi(\spinf M_\text{f}+\sqrt{\tilde{r}_\text{tide}^3/M_\text{f}})}\,,
\end{align}
where upper/lower signs hold for prograde/retrograde orbits.  So far
--- aside from the inclusion of $\spin$-dependent terms in
Eqs.\,(\ref{eq:xi-tide})-(\ref{eq:ftide}) and in obtaining $\mBHf$ and
$\spinf$ from the model of~\cite{Pannarale2012, Pannarale2014} ---
nothing differs from the approach laid out in Paper I.  We would like
to note that in the process of calculating $\spinf$ and $\mBHf$ one
must also determine another quantity that plays a role in the
gravitational waveform model.  This is the mass of the torus that may
remain around the \ac{BH} at late times, $\mTorus$, modelled using the
fitting formula~\cite{Foucart:2012nc}
\begin{align}
  \label{eq:diskMass}
  \f{\mTorus}{M_\text{b,NS}}=\f{0.296\tilde{r}_\text{tide}-0.171r_\text{ISCO}}{R_\text{NS}}\,,
\end{align}
where $M_\text{b,NS}$ is the rest-mass of the \ac{NS} in isolation and
$r_\text{ISCO}$ is the radius of the \ac{ISCO} of the initial \ac{BH}
in isolation~\cite{Bardeen:1972fi}:
\begin{align}
  \bar{r}_\text{ISCO} &= [3+Z_2\mp\sqrt{(3-Z_1)(3+Z_1+2Z_2)}]\,,\nn\\
  Z_1 &= 1 +
  (1-\spin^2)^{1/3}\left[(1+\spin)^{1/3}+(1-\spin)^{1/3}\right]\,,\nn\\
  \label{eqs:rISCO}
  Z_2 &= \sqrt{3\spin^2+Z_1^2}\,.
\end{align}

As in Paper I, we write the amplitude $A_\text{Phen}(f)$ of the
frequency-domain signal
$h_\text{Phen}(f)=A_\text{Phen}(f)e^{i\Phi_\text{Phen}(f)}$ as a sum
of three terms:
\begin{eqnarray}
  \label{eq:PhenoMixedAmp}
  A_\text{Phen}(f) &=&
  A_\text{PN}(f)w_{\epsilon_\text{ins}\tilde{f}_0,d+\sigma_\text{tide}}^- \nn\\
  &+& 1.25\gamma_1f^{5/6}w_{\tilde{f}_0,d+\sigma_\text{tide}}^- \nn\\
  &+& \mathcal{A}_\text{RD}(f)w_{\tilde{f}_0,d+\sigma_\text{tide}}^+\,,
\end{eqnarray}
where $A_\text{PN}(f)$ is the inspiral contribution, based on the
stationary-phase approximation and obtained by combining a
$3$\ac{PN}-order time-domain expansion of the amplitude and the
TaylorT4 description for the phase (see \cite{Santamaria:2010yb} for
further details); the second term models the premerger and merger
(strong-field) modifications to the \ac{PN} inspiral (the $\gamma_1$
coefficient is provided, once more, in \cite{Santamaria:2010yb}); and
$\mathcal{A}_\text{RD}$ is the ringdown amplitude.  This is modelled
via a Lorentzian $\mathcal{L}(f,f_0,\sigma)\equiv
\sigma^2/[(f-f_0)^2+\sigma^2/4]$:
\beq%
\mathcal{A}_\text{RD}(f)=\epsilon_\text{tide}\delta_1\mathcal{L}(f,
f_\text{RD}(\spinf,\mBHf),\delta_2^\prime
f_\text{RD}/\mathcal{Q}(\spinf))f^{-7/6}\,,%
\nn \\\label{eq:RDAmp}%
\eeq
where\footnote{The third argument here corrects a typo in Eq.\,(3) of
  Paper I.} $\delta_1$ is the ringdown amplitude fitted to \bbh hybrid
waveform data in~\cite{Santamaria:2010yb}, $\epsilon_\text{tide}$ is a
\nsbh correction discussed later on in this section, and
$\delta_2^\prime$ is a fudge factor which accounts for errors in the
model used to compute $\spinf$, as this spin parameter is in turn used
to determine the quality factor $\mathcal{Q}$ of the \ac{BH} remnant
(once again via the fitting formulas of~\cite{BertiCardosoWill}).  We
note that the PhenomC model also uses a fudge factor, $\delta_2$, but
$\spinf$ is determined using the formulas
in~\cite{Rezzolla-etal-2007b}. Therefore, rather than correcting the
PhenomC parameter $\delta_2$ as we did in Paper I, in this paper we
introduce a $\delta_2^\prime$ parameter in order to disentangle
PhenomC and our model more clearly.  The remaining elements of
Eq.\,(\ref{eq:PhenoMixedAmp}) to be discussed are the windowing
functions.  As for the PhenomC model, $d$ is set to $0.015$, but, just
as in Paper I, we allow for a correction $\sigma_\text{tide}$ to the
width, we do not necessarily tie the central windowing frequencies to
the \ac{BH} remnant ringdown frequency, and we do not always fix the
central windowing frequencies of the first two terms in
$A_\text{Phen}(f)$ --- $\epsilon_\text{ins}\tilde{f}_0$ and
$\tilde{f}_0$, respectively --- to the same value.  More details on
how $\sigma_\text{tide}$, $\epsilon_\text{ins}$ and $\tilde{f}_0$ are
determined are given further on in this section.

To summarize, in order to build $A_\text{Phen}(f)$ given $\mBH$,
$\spin$, $\mNS$ and an \ac{EOS} (which determines the \ac{NS} radius
$\rNS$ and its baryonic rest-mass $M_\text{b,NS}$), one must begin by
computing: (1) $\gamma_1$ and $\delta_1$ according
to~\cite{Santamaria:2010yb}, (2) $\mTorus$ using
Eqs.\,(\ref{eq:diskMass})-(\ref{eqs:rISCO}), (3) $\mBHf$ and $\spinf$
following the model reported in~\cite{Pannarale2012,Pannarale2014},
(4) $\fRD(\mBHf,\spinf)$ and $\mathcal{Q}(\spinf)$ via the fits
of~\cite{BertiCardosoWill}, and (5) $\fTide$, following
Eqs.\,(\ref{eq:xi-tide})-(\ref{eq:rtide}).  At this point the model
splits into four cases, depending on the values of $\fTide$, $\fRD$,
and $\mTorus$.  These cases reflect the different phenomenology of
\nsbh binary mergers observed in the $59$ simulations used to build
the model (see Sec.\,\ref{sec:NRdata}): ``disruptive,''
``nondisruptive,'' and ``mildly disruptive'' with and without a torus
remnant.

Before laying out the necessary details about the four alternative
waveforms in the remaining subsections, we will briefly explain how
the four phenomenological models were obtained.  Each of the hybrid
waveform amplitudes corresponding to nondisruptive mergers --- out of
the $59$ hybrids used to build our model --- was fitted with the
ansatz in Eq.\,(\ref{eq:PhenoMixedAmpND}) below, leaving
$\epsilon_\text{tide}$, $\sigma_\text{tide}$, and $\delta_2^\prime$ as
free coefficients.  The values of the free coefficients were then
themselves fitted as detailed in
Eqs.\,(\ref{eq:epsTideND})-(\ref{eq:delta2prime}).  The same procedure
was followed for the disruptive mergers, where, this time, the
waveform amplitude ansatz is given in Eq.\,(\ref{eq:PhenoMixedAmpD})
and the values of the free coefficients $\epsilon_\text{ins}$ and
$\sigma_\text{tide}$ are fitted with
Eqs.\,(\ref{eq:epsIns})-(\ref{eq:xDprime}).  All equations involved in
this process were inspired by the nonspinning study reported in Paper
I.  For the two mildly disruptive cases, we adopted a strategy that is
similar to the one discussed in Paper I: we picked and combined
ingredients from disruptive and nondisruptive phenomenological
waveforms, without having to perform additional fits.

\subsubsection{Nondisruptive Mergers}\label{sec:ND}
\begin{figure}[!tb]
  \includegraphics[width=\columnwidth,clip=true]{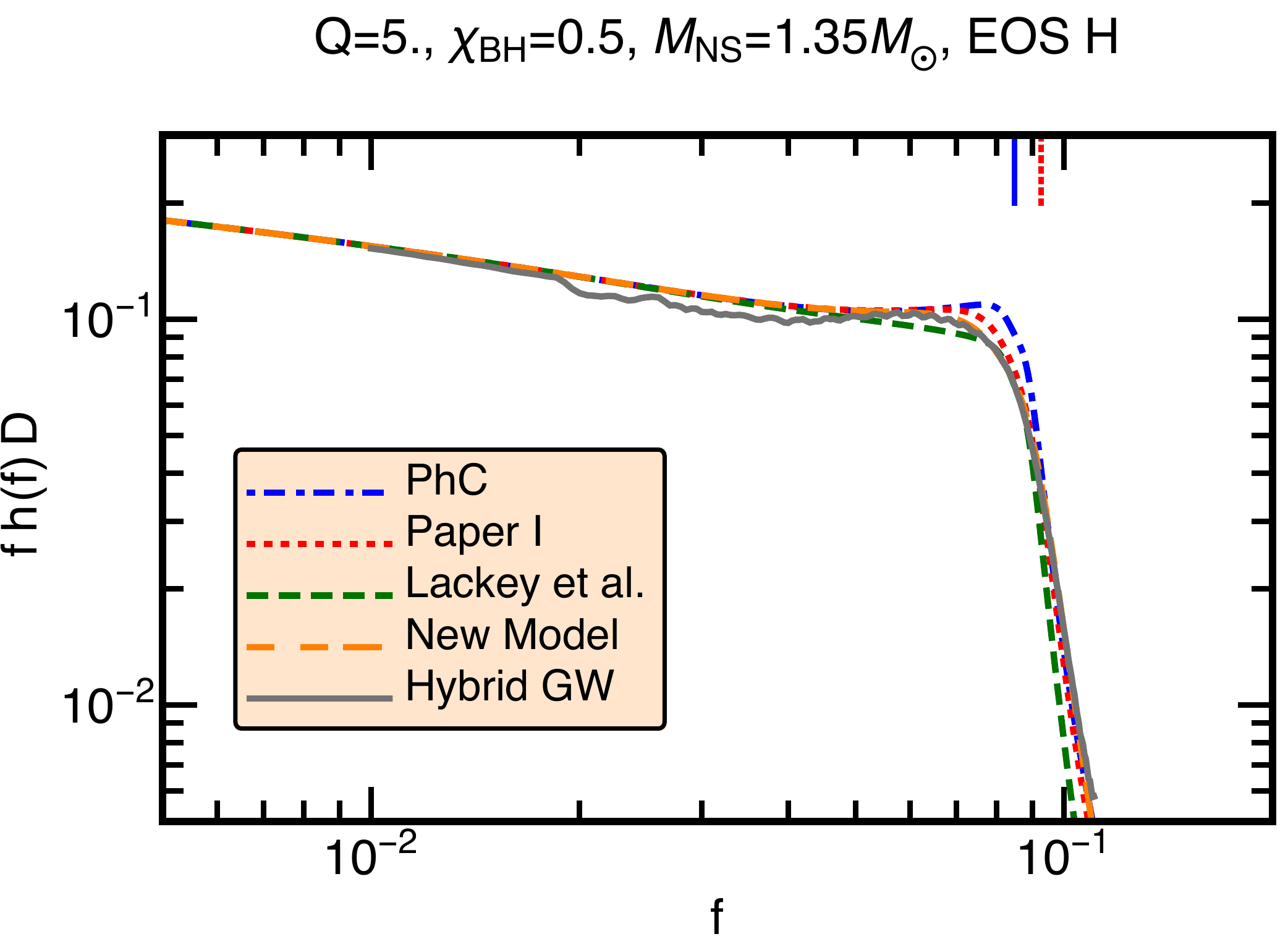} \\
  \includegraphics[width=\columnwidth,clip=true]{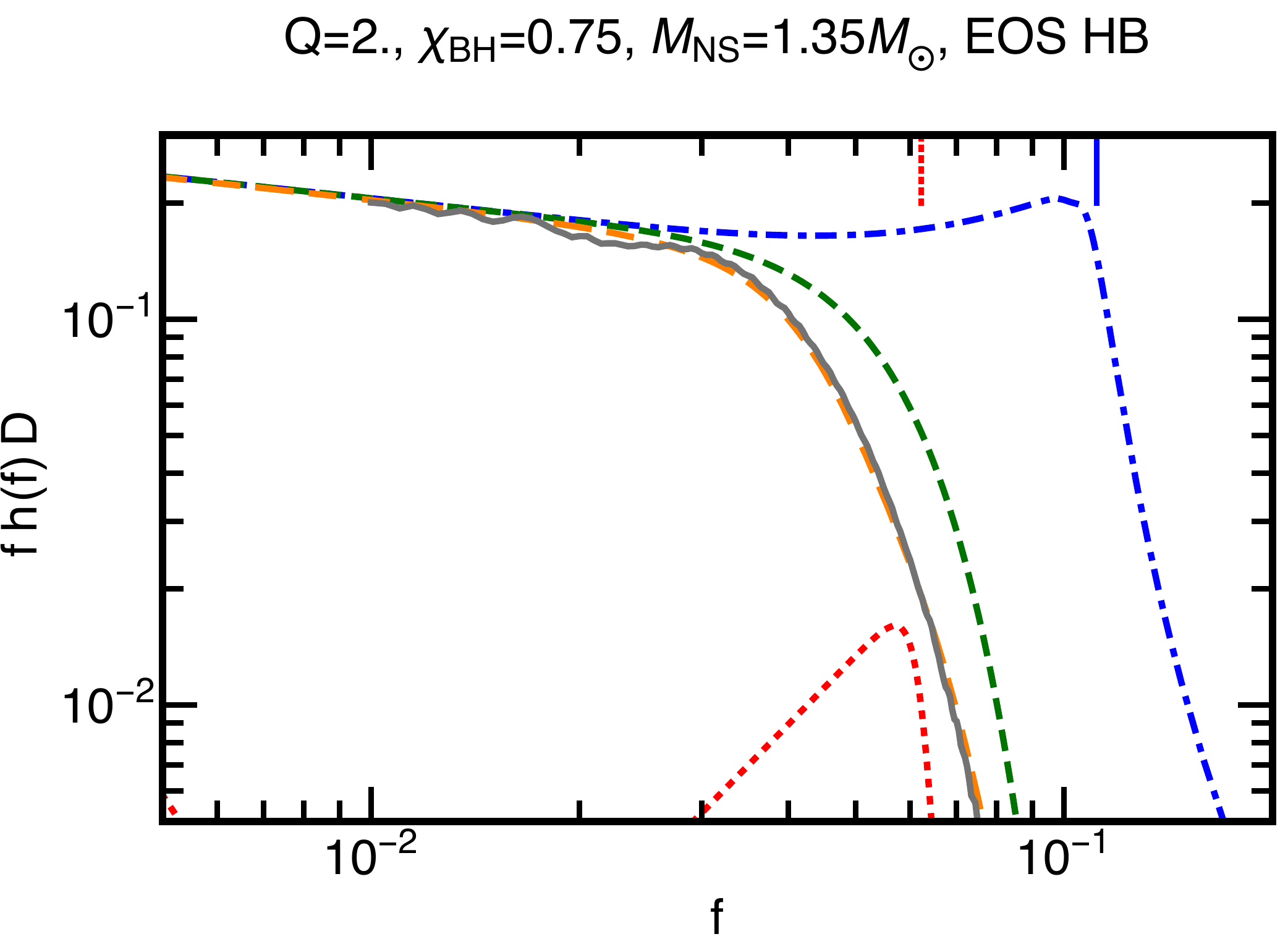} \\
  \includegraphics[width=\columnwidth,clip=true]{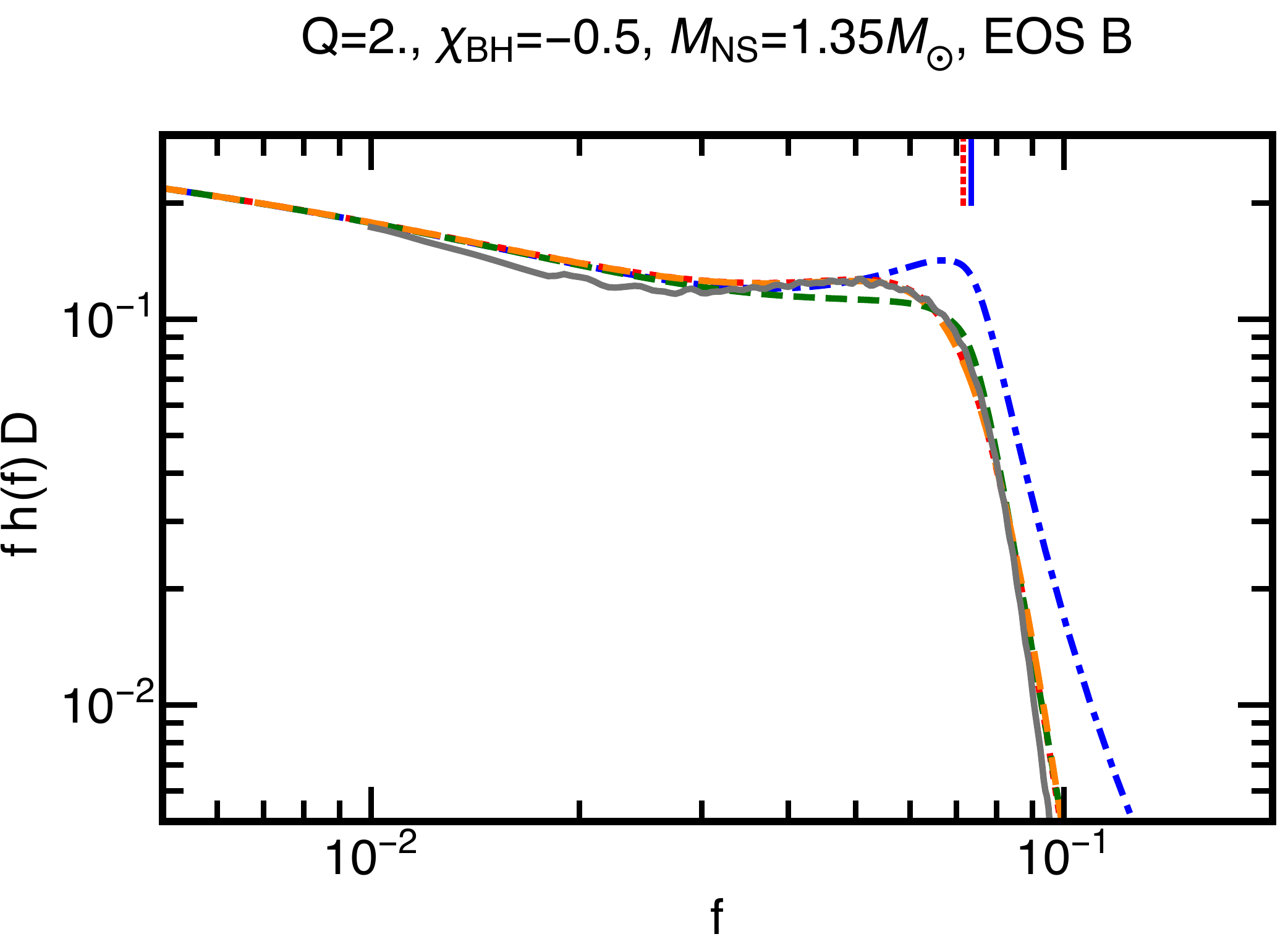}
  \caption{Examples of nondisruptive, disruptive, and mildly
    disruptive merger waveforms (from top to bottom).  The hybrid
    waveform (in gray) is compared to the PhenomC model (dot-dashed
    blue), to the nonspinning \nsbh model of Paper I (dotted red), to
    the \nsbh model of~\cite{Lackey2014} (dashed green), and to the
    model developed in this paper (long-dashed orange).  The short
    vertical lines denote $\fRD$ and $\fTide$ in blue and dotted red,
    respectively. \label{fig:GWexamplePlots}}
\end{figure}

If $\fTide\geq\fRD$ and $\mTorus=0$, the merger is ``nondisruptive.''
Notice that the first condition differs slightly from the one used in
the nonspinning waveform model reported in Paper I. Additionally, we
had not explicitly required the second condition in Paper I, because
we never encountered nonspinning cases for which the analytical models
would predict $\mTorus>0$ and $\fTide>\tfRD$.  For nondisruptive
mergers the binary can complete its full inspiral stage, and therefore
we set $\epsilon_\text{ins}=1$ in Eq.\,(\ref{eq:PhenoMixedAmp}).
Moreover, the \ac{BH} remnant ringdown is excited, so the windowing
functions in Eq.\,(\ref{eq:PhenoMixedAmp}) can be centered around
$\tilde{f}_0=\tfRD$\footnote{Nondisruptive \nsbh mergers essentially
  behave as \bbh mergers. In PhenomC waveforms, $\tilde{f}_0$ is set
  to $0.98\fRD$; this quantity in turn depends on the mass of the
  \ac{BH} remnant, which is set to the sum of the individual masses of
  the binary components, $m_0$.  In our model the mass of the \ac{BH}
  remnant is instead determined with the approach described
  in~\cite{Pannarale2012}.  More specifically, $\mBHf$ is always
  smaller than $m_0$, the final mass choice made in PhenomC.  This
  increases $\fRD$, hence the need for the extra $0.99$ in our
  definition of $\tilde{f}_0$.}.  $A_\text{Phen}(f)$ thus takes the
form
\beq%
\label{eq:PhenoMixedAmpND}
A_\text{Phen}(f) &=&
A_\text{PN}(f)w_{\tfRD,d+\sigma_\text{tide}}^- \nn\\
&+& 1.25\gamma_1f^{5/6}w_{\tfRD,d+\sigma_\text{tide}}^- \nn\\
&+& \mathcal{A}_\text{RD}(f)w_{\tfRD,d+\sigma_\text{tide}}^+\,,
\eeq%
where $\mathcal{A}_\text{RD}$ is given by Eq.\,(\ref{eq:RDAmp}).

The parameters entering Eqs.\,(\ref{eq:RDAmp}) and
(\ref{eq:PhenoMixedAmpND}) are given by:
\begin{align}
\label{eq:epsTideND}
  \epsilon_\text{tide} &= 2w_{x_1,d_1}^+(x_\text{ND})\,,%
\end{align}%
where $x_1=-0.0796251$, $d_1= 0.0801192$, and%
\begin{align}
  x_\text{ND} &\equiv \left(\f{\fTide - \tfRD}{\tfRD}\right)^2 - 0.571505\mathcal{C}\nn\\
  &- 0.00508451\spin\,;%
\end{align}
\be%
\label{eq:sigmaTideND}%
\sigma_\text{tide} = 2w_{x_2,d_2}^-(x_\text{ND}^\prime)\,,%
\ee%
where $x_2=-0.206465$, $d_2=0.226844$ and
\begin{align}
  x_\text{ND}^\prime &\equiv \left(\f{\fTide - \tfRD}{\tfRD}\right)^2 - 0.657424\mathcal{C}\nn\\
  &- 0.0259977\spin\,;%
\end{align}
and finally
\be%
\label{eq:delta2prime}
\delta_2^\prime = Aw_{x_3,d_3}^-\left(\f{\fTide -
    \tfRD}{\tfRD}\right)\,,%
\ee%
where $A=1.62496$, $x_3=0.0188092$ and $d_3=0.338737$.
Eq.\,(\ref{eq:epsTideND}) slowly suppresses the ringdown of the
\ac{BH} remnant as the merger becomes less and less similar to the
\bbh case.  We notice that, at variance with the nonspinning
formulation of Paper I, we had to introduce two different independent
variables for the two waveform parameters $\epsilon_\text{tide}$ and
$\sigma_\text{tide}$, i.e.~$x_\text{ND}\neq x_\text{ND}^\prime$.
Furthermore, the two independent variables now contain terms linear in
$\spin$.  An example of nondisruptive merger spectrum is shown in the
first panel in Figure \ref{fig:GWexamplePlots}.  Here, the hybrid
waveform (in gray) is compared to the PhenomC model (dot-dashed blue),
to the nonspinning \nsbh model of Paper I (dotted red), to the \nsbh
model of~\cite{Lackey2014} (dashed green), and to the model developed
in this paper (long-dashed orange).

\subsubsection{Disruptive Mergers}\label{sec:D}
If $\fTide<\fRD$ and $\mTorus>0$, then the merger is ``disruptive,''
the \ac{NS} material is scattered around the \ac{BH}, and the ringdown
contribution to Eq.\,(\ref{eq:PhenoMixedAmp}) vanishes,
i.e.~$\epsilon_\text{tide}=0$ in Eq.\,(\ref{eq:RDAmp}).  As in the
case of nondisruptive mergers, the definition of this class is
slightly different from its corresponding definition in Paper I.  For
disruptive mergers the waveform model reads:
\beq
\label{eq:PhenoMixedAmpD}%
A_\text{Phen}(f) &=&
A_\text{PN}(f)w_{\epsilon_\text{ins}\fTide,d+\sigma_\text{tide}}^- \nn\\
&+& 1.25\gamma_1f^{5/6}w_{\fTide,d+\sigma_\text{tide}}^-\,,%
\eeq%
which is equivalent to Eq.\,(\ref{eq:PhenoMixedAmp}) with
$\tilde{f}_0=\fTide$ and $\epsilon_\text{tide}=0$.  The remaining
parameters to be prescribed are $\epsilon_\text{ins}$ and
$\sigma_\text{tide}$.  These are given by:
\be%
\label{eq:epsIns}
\epsilon_\text{ins} = a_1 + b_1x_\text{D}\,,%
\ee
where $a_1=1.29971$, $b_1=-1.61724$ and
\begin{align}%
  x_\text{D} &\equiv \f{M_\text{b,torus}}{M_\text{b,NS}} +
  0.424912\mathcal{C} + 0.363604\sqrt{\nu} \nn\\
  &-0.0605591\spin\,,%
\end{align}
$\nu=\mNS\mBH/m_0^2$ being the symmetric mass ratio; and
\be%
\label{eq:sigmaTideD}%
\sigma_\text{tide} = a_2 + b_2x_\text{D}^\prime\,,%
\ee
where $a_2=0.137722$, $b_2=-0.293237$ and
\begin{align}
  \label{eq:xDprime}
  x_\text{D}^\prime &\equiv \f{M_\text{b,torus}}{M_\text{b,NS}} -
  0.132754\mathcal{C} + 0.576669\sqrt{\nu} \nn\\
  &- 0.0603749\spin
  -0.0601185\spin^2 \nn\\
  &- 0.0729134\spin^3\,.%
\end{align}
As in the case of nondisruptive mergers, two distinct definitions of
the independent variable are used for the two waveform parameters,
which was not the case in Paper I.  Incidentally, this allows us to
simplify the functional form of the parameter $\sigma_\text{tide}$
with respect to Paper I.  An example of disruptive merger spectrum is
shown in the middle panel of Figure \ref{fig:GWexamplePlots}.  Notice
that the nonspinning model of Paper I is completely inaccurate when
high spin and strong tidal effects come into play, as the model did
not account for the combination of the two.

\subsubsection{Mildly Disruptive Mergers with no Torus
  Remnant}\label{sec:MD-woTorus}
If $\fTide<\fRD$ and $\mTorus=0$, then the merger is ``mildly
disruptive,'' but \emph{no} remnant torus is formed during the
coalescence.  The behavior of the \ac{GW} amplitude for this class of
mergers is captured by using the disruptive merger waveform model
(Sec.\,\ref{sec:D}) and modifying two of its features.  The first
modification concerns the central frequencies of the windowing
functions entering Eq.\,(\ref{eq:PhenoMixedAmpD}).  In the present
case we set
\beq%
\label{eq:PhenoMixedAmpMDNoDisk}%
A_\text{Phen}(f) &=&
A_\text{PN}(f)w_{f_1,d+\sigma_\text{tide}}^- \nn\\
&+& 1.25\gamma_1f^{5/6}w_{f_2,d+\sigma_\text{tide}}^-\,,%
\eeq
where $f_1=(1-Q^{-1})\tfRD+Q^{-1}\epsilon_\text{ins}\fTide$ and
$f_2=(1-Q^{-1})\tfRD+Q^{-1}\fTide$ (our definition of the binary mass
ratio is such that $Q\geq 1$). These two frequencies are introduced in
order to obtain a smooth transition from nondisruptive to disruptive
merger waveforms, which use $\tfRD$, and $\epsilon_\text{ins}\fTide$
and $\fTide$ as windowing central frequencies, respectively.  This
constitutes a major improvement with respect to Paper I, where the low
number of available nonspinning mildly disruptive simulations had not
allowed us to go into such fine modelling details.  Additionally, the
scaling with $Q$ is such that the model can accurately reproduce the
data for mildly disruptive mergers with no torus remnant.

The second modification with respect to the disruptive merger waveform
model in Eq.\,(\ref{eq:PhenoMixedAmpD}) is in the calculation of
$\sigma_\text{tide}$. For mildly disruptive mergers with no torus
remnant, this calculation is performed by averaging the disruptive and
the nondisruptive prescriptions: i.e., we evaluate
Eqs.\,(\ref{eq:sigmaTideND}) and (\ref{eq:sigmaTideD}) and divide
their sum by two.  As for $f_1$ and $f_2$, this modification allows
for a smooth transition from disruptive to nondisuptive merger
waveforms, and it is a notable improvement with respect to Paper I.

\subsubsection{Mildly Disruptive Mergers with a Torus
  Remnant}\label{sec:MD-wTorus}
\begin{figure*}[!tb]
  \begin{tabular*}{\textwidth}{c@{\extracolsep{\fill}}c}
    \includegraphics[width=\columnwidth,clip=true]{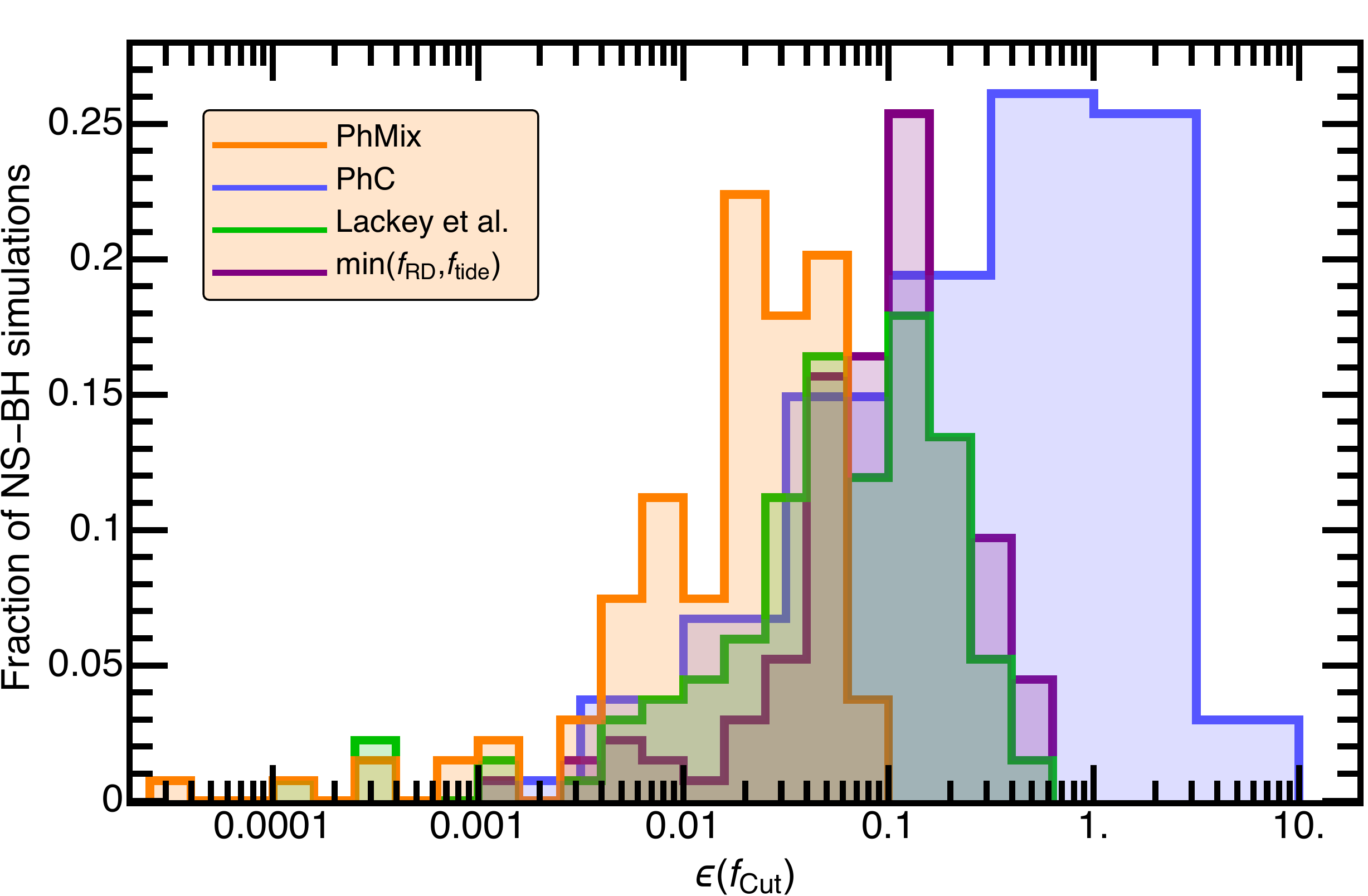}
    &
    \includegraphics[width=\columnwidth,clip=true]{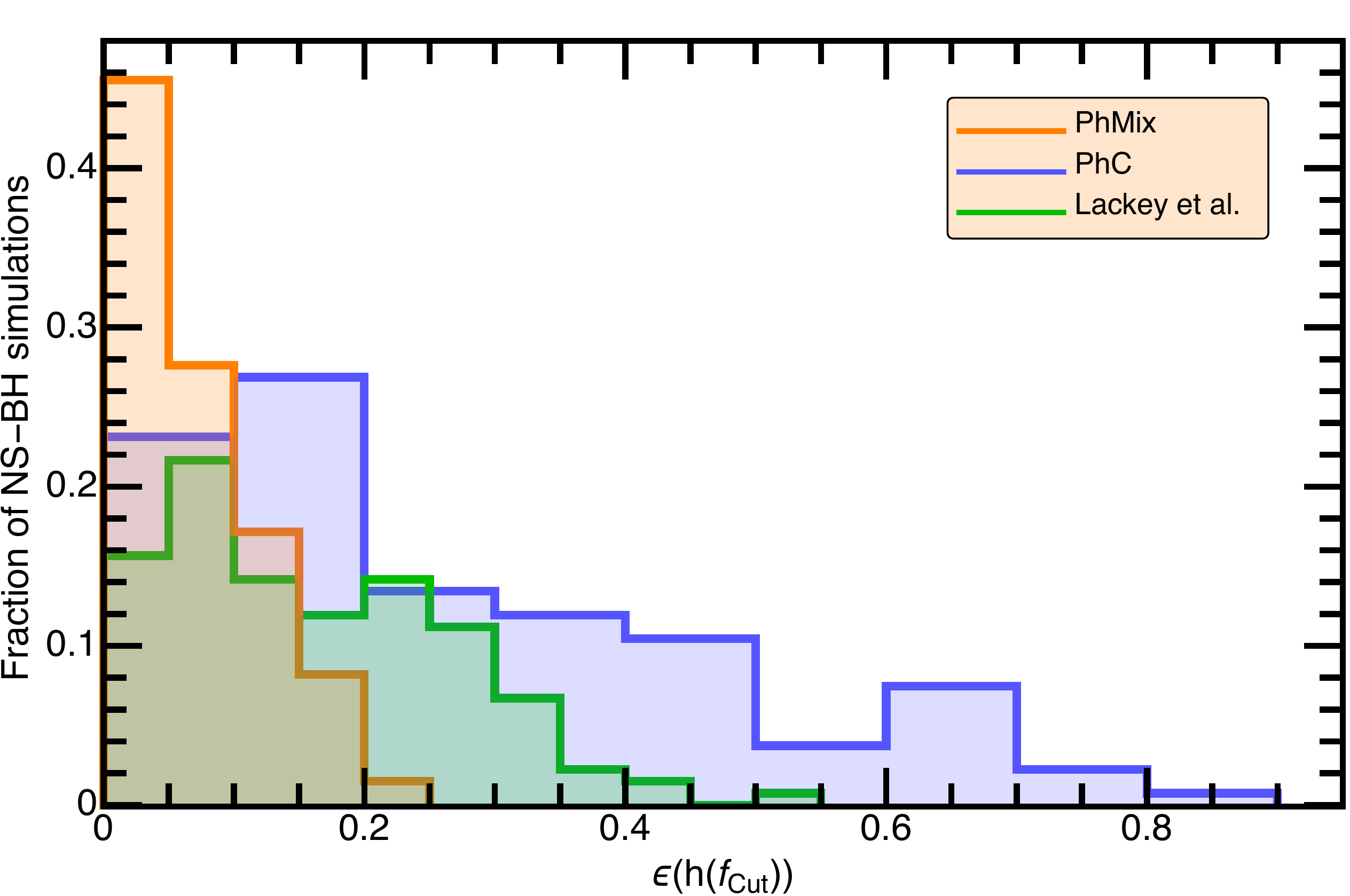}
  \end{tabular*}
  \caption{Relative error distribution over the $134$ \nsbh merger
    simulations for $\fCut$ (left panel) and $h(\fCut)$ (right panel).
    All $134$ results from the hybrid waveforms are compared to the
    predictions obtained from different models: the waveform model
    reported in this paper (``PhMix,'' orange), PhenomC (``PhC,''
    blue), the waveform model of~\cite{Lackey2014} (``Lackey et al.,''
    green), and the simple estimate given by $\min(\fRD,\fTide)$
    (purple).  As discussed in the text this last proxy does not
    provide a waveform amplitude model, so it does not appear in the
    panel on the right.  \label{fig:fCutAll}}
\end{figure*}

If $\fTide\geq\fRD$ and $\mTorus>0$, then the merger is ``mildly
disruptive'' and a small remnant torus is produced.  Mathematically,
this class of mergers arises from a shortcoming of the approximations
that lead to $\fTide$, $\mTorus$, and, possibly, $\fRD$: one would not
expect a remnant torus to be formed if the system cannot reach the
onset of tidal disruption.  Physically, what the combination
$\fTide\geq\fRD$ and $\mTorus>0$ suggests (and what the hybrid
waveforms that fall in these categories indeed show) is that some
mildly disruptive mergers can achieve both a \ac{QNM} excitation of
the \ac{BH} remnant and the formation of a small remnant accretion
torus\footnote{$\mTorus$ was found to be $\leq 0.01\mSun$ for these
  tuning cases: this is clearly below the precision of
  Eq.\,(\ref{eq:diskMass}).}.  From a phenomenological point of view
this is to be expected, as the outer layers of the \ac{NS} may be
stripped to form the torus, while the core of the \ac{NS} (or a
fraction of it) may accrete onto the \ac{BH} coherently enough to
trigger the \ac{BH} \ac{QNM} oscillations (with regards to this topic,
see Fig.\,17 in \cite{Kyutoku:2011vz} and our discussion of
Fig.\,\ref{fig:fCutContours_spins} below).  We wish to remark that
this level of sophistication in cataloguing mildly disruptive mergers
was not possible in Paper I, and is dictated by the fact that in the
present work we have a larger catalog of numerical-relativity data to
reproduce.

For mergers with $\fTide\geq\fRD$ and $\mTorus>0$, the waveform
amplitude model reads
\beq%
\label{eq:PhenoMixedAmpMDWithDisk}%
A_\text{Phen}(f) &=&
A_\text{PN}(f)w_{\epsilon_\text{ins}\tfRD,d+\sigma_\text{tide}}^- \nn\\
&+& 1.25\gamma_1f^{5/6}w_{\epsilon_\text{ins}\tfRD,d+\sigma_\text{tide}}^- \nn\\
&+& \mathcal{A}_\text{RD}(f)w_{\tfRD,d+\sigma_\text{tide}}^+\,,%
\eeq
where $\epsilon_\text{ins}$ is set according to
Eq.\,(\ref{eq:epsIns}), as for disruptive mergers, and
$\sigma_\text{tide}$ is given by Eq.\,(\ref{eq:sigmaTideND}), as for
nondisruptive mergers.  Notice that
Eq.\,(\ref{eq:PhenoMixedAmpMDWithDisk}) resembles
Eq.\,(\ref{eq:PhenoMixedAmpND}), with the exception of the use of
$\epsilon_\text{ins}$ in setting the first two windowing central
frequencies: this allows for a smooth transition between nondisruptive
and disruptive waveform models whenever the \ac{NS} is tidally
disrupted, but the excitation of the \ac{BH} remnant ringdown takes
place and no torus remnant is expected to survive the merger.  An
example of disruptive merger spectrum is shown in the last panel in
Figure \ref{fig:GWexamplePlots}; in this specific case,
Eq.\,(\ref{eq:PhenoMixedAmpMDWithDisk}) is used to model the waveform
amplitude.

\section{Testing the Waveform Model}\label{sec:tests}
Once the phenomenological \nsbh waveform amplitude model is formulated
and calibrated via fits to hybrid waveforms built upon
numerical-simulation data, its accuracy must be tested and
demonstrated.

The qualitative behavior of the spinning waveform model is the same as
for the nonspinning model formulated in Paper I. For the sake of
brevity, we do not show all $134$ \ac{GW} spectra.  We perform,
instead, a quantitative assessment of the accuracy of the model in
terms of two characteristic frequencies of each \ac{GW} spectrum ---
$f_{\rm Max}$ and $\fCut$ --- and the amplitude of each spectrum at
these two frequencies.  For the cutoff frequency $\fCut$, we adopt the
same, general definition introduced in Paper I, which also involves
introducing and determining $f_{\rm Max}$.  This definition has the
advantage of allowing for a straightforward comparison among \ac{GW}
spectra originating from different models and/or calculations for the
same binary, and for consistent comparisons among binaries with
different physical parameters.  The definition is as follows: $f_{\rm
  Max}$ is the frequency such that $f^2h(f)$ has a maximum, and
$\fCut$ is the frequency (greater than $f_{\rm Max}$) at which the
amplitude drops by one $e$-fold, i.e.,
\be%
\label{eq:fCut}%
e \fCut h(\fCut) = f_{\rm Max}h(f_{\rm Max})\,.%
\ee%
We stress once more that this definition of the cutoff frequency is
independent of the details of the waveform, and it works for any
$h(f)$ (given in either analytical or numerical form).

\begin{figure*}[!tb]
  \begin{tabular*}{\textwidth}{c@{\extracolsep{\fill}}c}
    \includegraphics[width=\columnwidth,clip=true]{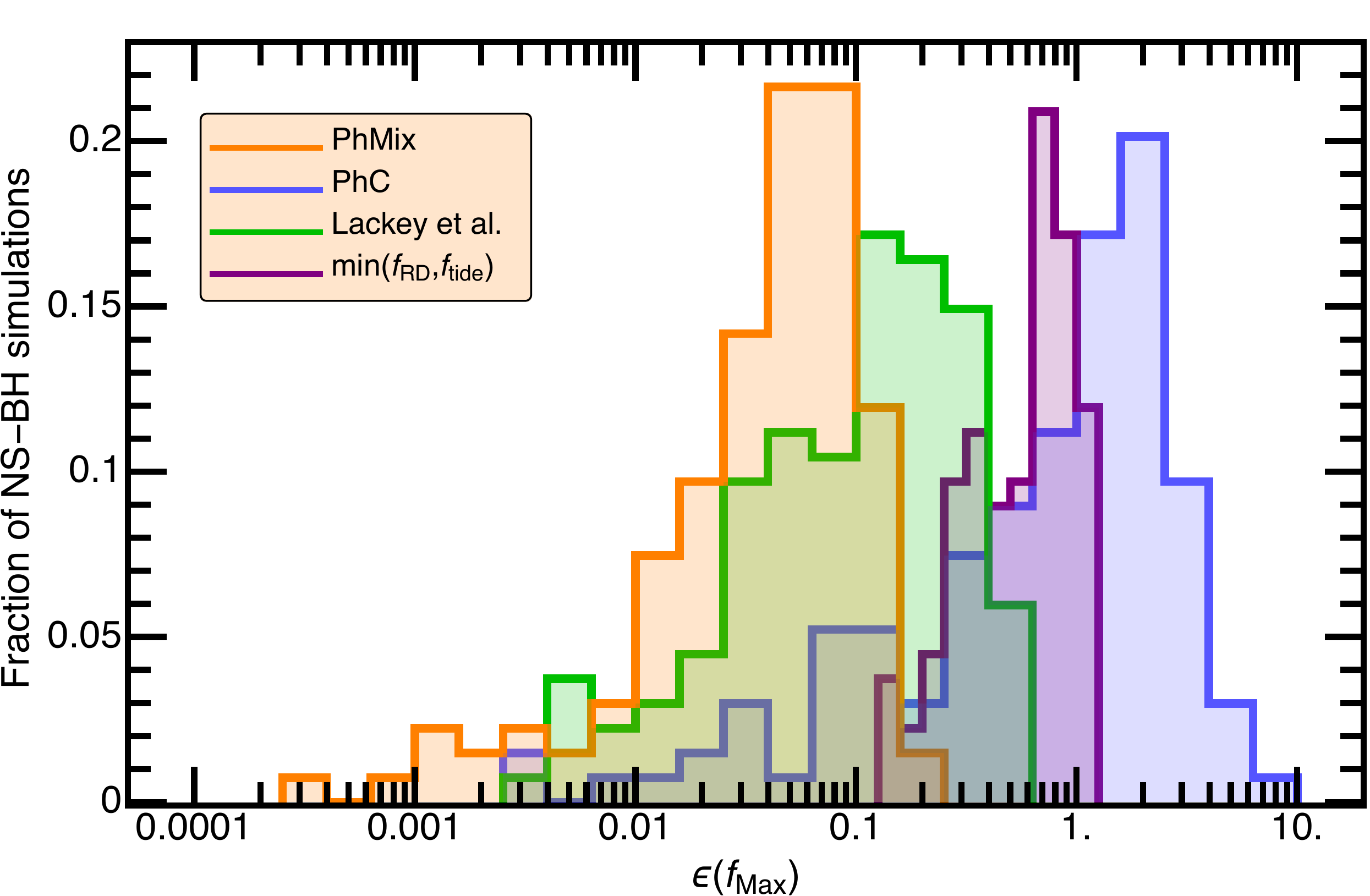}
    &
    \includegraphics[width=\columnwidth,clip=true]{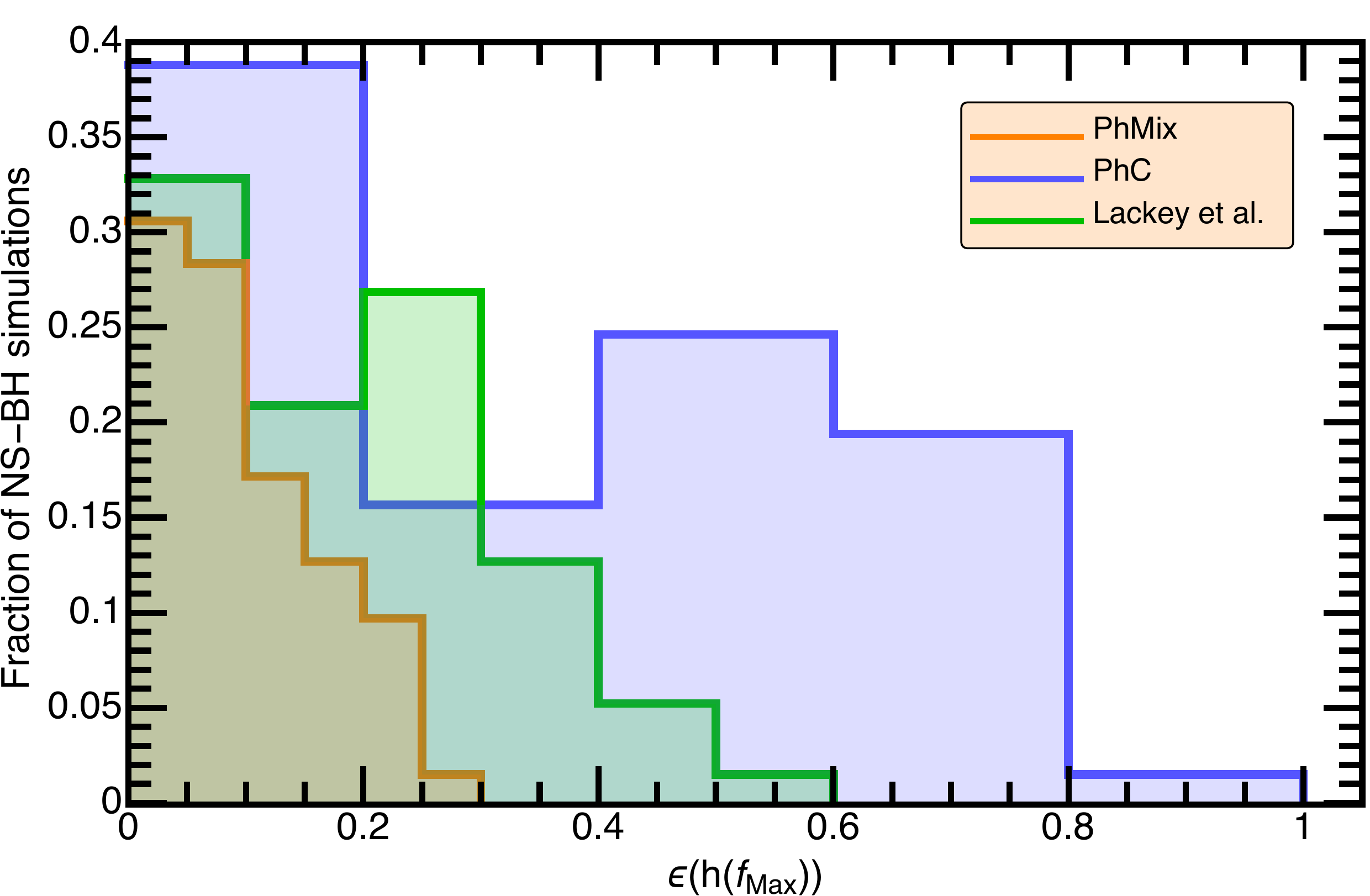}
  \end{tabular*}
  \caption{Relative error distribution over the $134$ \nsbh hybrid
    waveforms for $f_{\rm Max}$ (left panel) and $h(f_{\rm Max})$
    (right panel). Colors and labelling are the same as in Figure
    \ref{fig:fCutAll}. \label{fig:fMaxAll}}
\end{figure*}

Figures \ref{fig:fCutAll} and \ref{fig:fMaxAll} show the distributions
of the relative errors ($\epsilon$) on $\fCut$ and $h(\fCut)$, and
$f_{\rm Max}$ and $h(f_{\rm Max})$, respectively.  For each of the
$134$ hybrid waveforms, we determine these four quantities and compare
them to the values predicted for them by several models.  These are:
(1) the waveform model reported in this paper (labelled ``PhMix'' in
the Figures), (2) PhenomC (``PhC''), (3) the waveform model
of~\cite{Lackey2014} (``Lackey et al.,''), and (4) the simple
prescription $\min(\fRD, \fTide)$, which, of course, cannot be used to
predict $h(\fCut)$ and $h(f_{\rm Max})$, but just as a tentative proxy
for $\fCut$ and $f_{\rm Max}$.  While the low frequency, inspiral
regime is described by construction in the same way by all models, our
new phenomenological model clearly introduces a considerable
improvement in terms of accurately predicting all four high-frequency
features of \acp{GW} emitted by spinning \nsbh coalescing binaries.
Both the mean and maximum values of the $\epsilon(\fCut)$ and
$\epsilon(f_{\rm Max})$ distributions are considerably reduced when
using our model.  The maximum relative error on the cutoff frequency
is of order $\sim 10$\%, to be compared with $\sim 60$\% for the model
of~\cite{Lackey2014} and the proxy $\min(\fRD, \fTide)$, and to even
higher relative errors for PhenomC, which was not designed for \nsbh
binaries.  The simple prescription $\min(\fRD, \fTide)$ is a better
proxy for $\fCut$ than it is for $f_{\rm Max}$.  The advantages of our
model are equally striking when considering $\epsilon(h(\fCut))$ and
$\epsilon(h(f_{\rm Max}))$.  Both distributions are peaked at $<0.05$
and fall off rapidly for the phenomenological \nsbh model, whereas
this is not the case for PhenomC and the model of~\cite{Lackey2014}.
In conclusion, our approach is more accurate than all of these
existing alternatives in modelling the \ac{GW} spectra of \nsbh
binaries.

\begin{figure*}[!tb]
  \begin{tabular*}{\textwidth}{c@{\extracolsep{\fill}}c}
    \includegraphics[width=\columnwidth,clip=true]{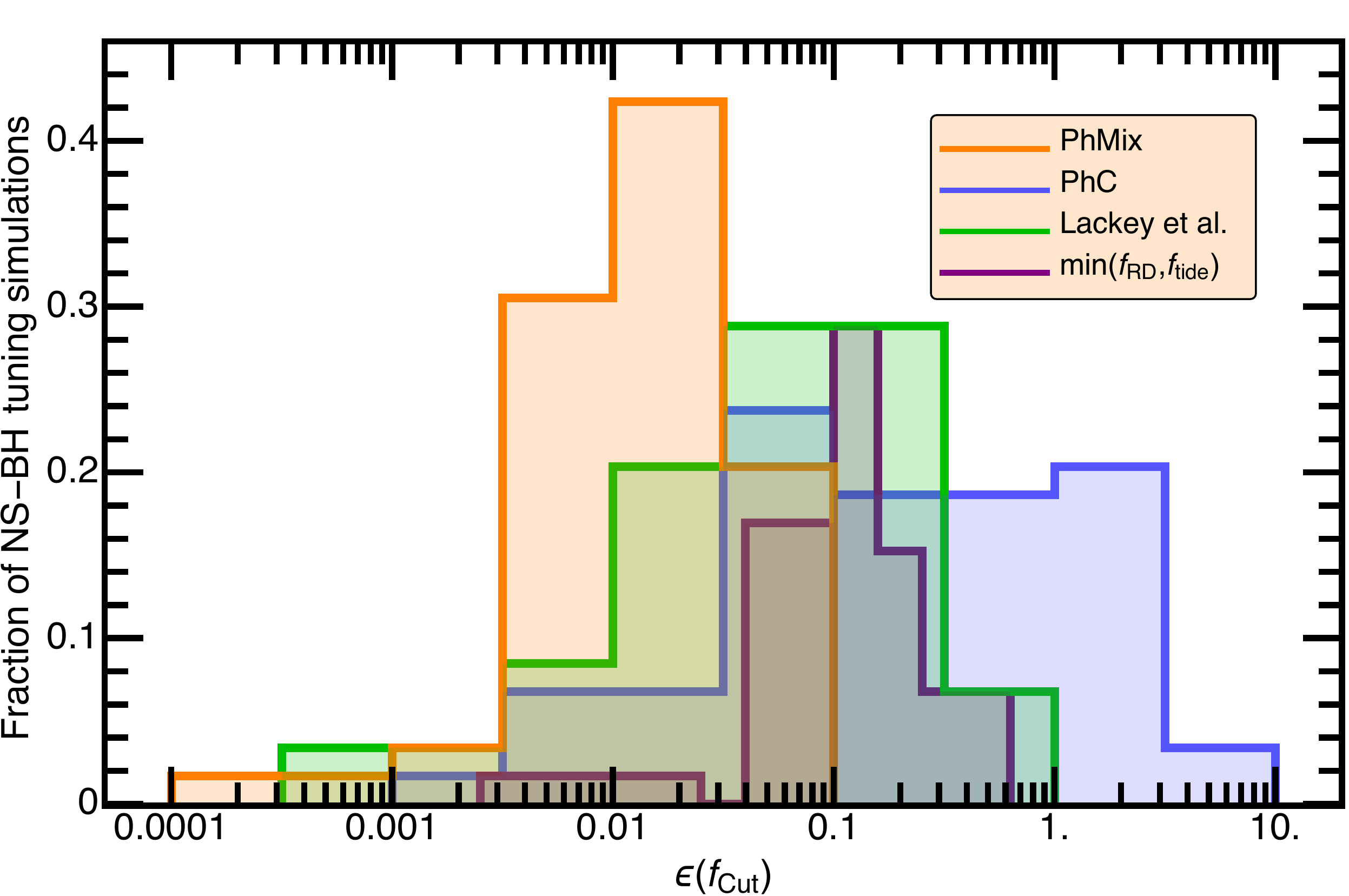}
    &
    \includegraphics[width=\columnwidth,clip=true]{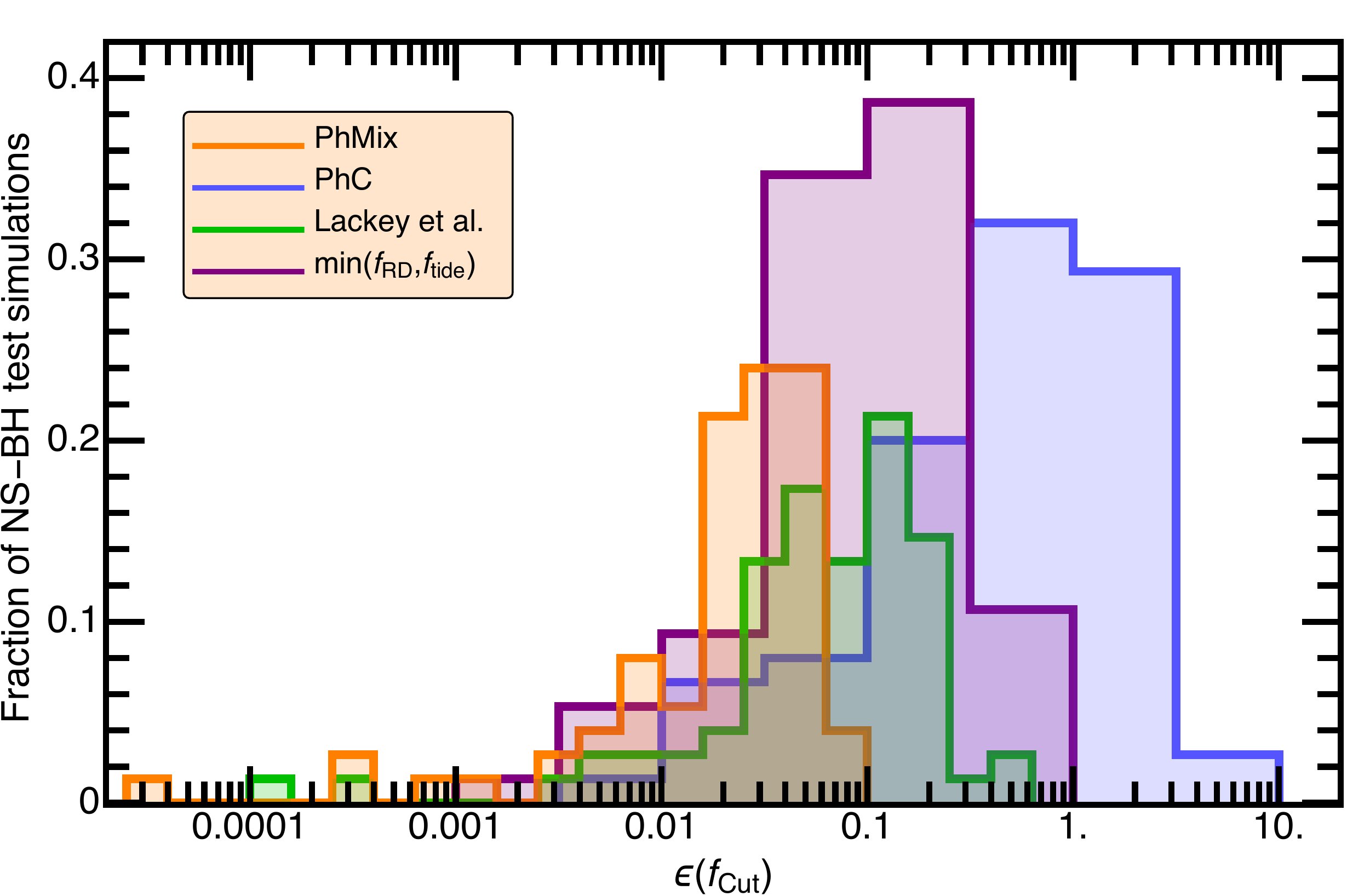}
  \end{tabular*}
  \caption{Relative error distribution on $\fCut$ (as in Figure
    \ref{fig:fCutAll}), but now the left panel refers to the $59$
    ``calibration binaries'' used to build the model, while the right
    panel refers to the $75$ ``test binaries'' that were not used to
    build the model. \label{fig:fCut}}
\end{figure*}

In Figure \ref{fig:fCut} we separate the two contributions to the
$\epsilon(\fCut)$ distribution shown in Figure \ref{fig:fCutAll}: in
the left panel we show the distribution of the relative error on the
cutoff frequency for the $59$ binaries use to {\em calibrate} the
model, while in the right panel we report the same distribution for
the $75$ binaries that were only used to {\em test} the model.  A
comparison with the left panel of Figure \ref{fig:fCutAll} shows that
the performance of our model is not dominated by results for
``calibration binaries.''  Analogous behaviors were obtained for
$\epsilon(f_{\rm Max})$, $\epsilon(h(\fCut))$ and $\epsilon(h(f_{\rm
  Max}))$.

\begin{figure*}[!tb]
  \begin{tabular*}{\textwidth}{c@{\extracolsep{\fill}}c}
    \includegraphics[width=\columnwidth,clip=true]{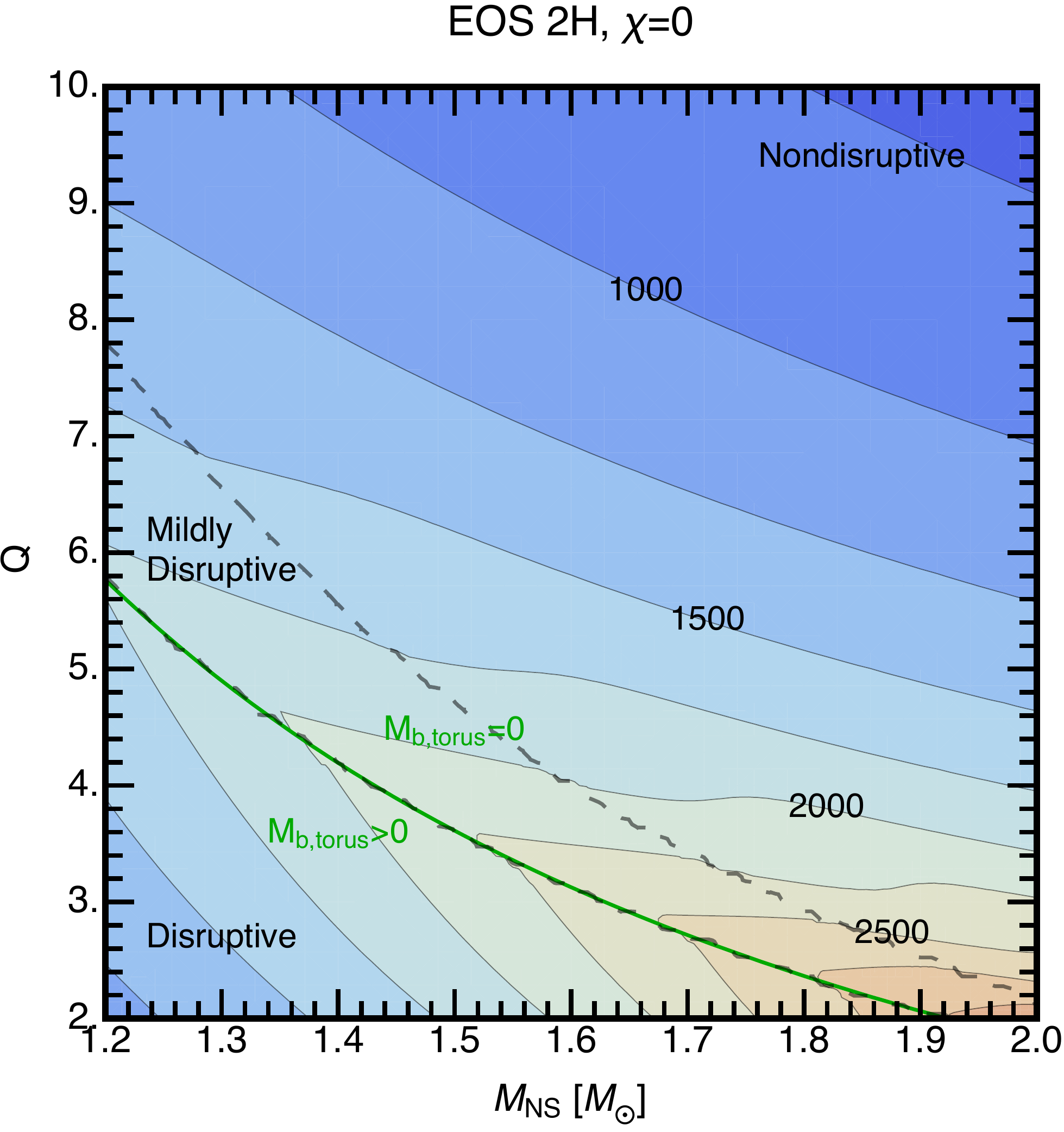}
    &
    \includegraphics[width=\columnwidth,clip=true]{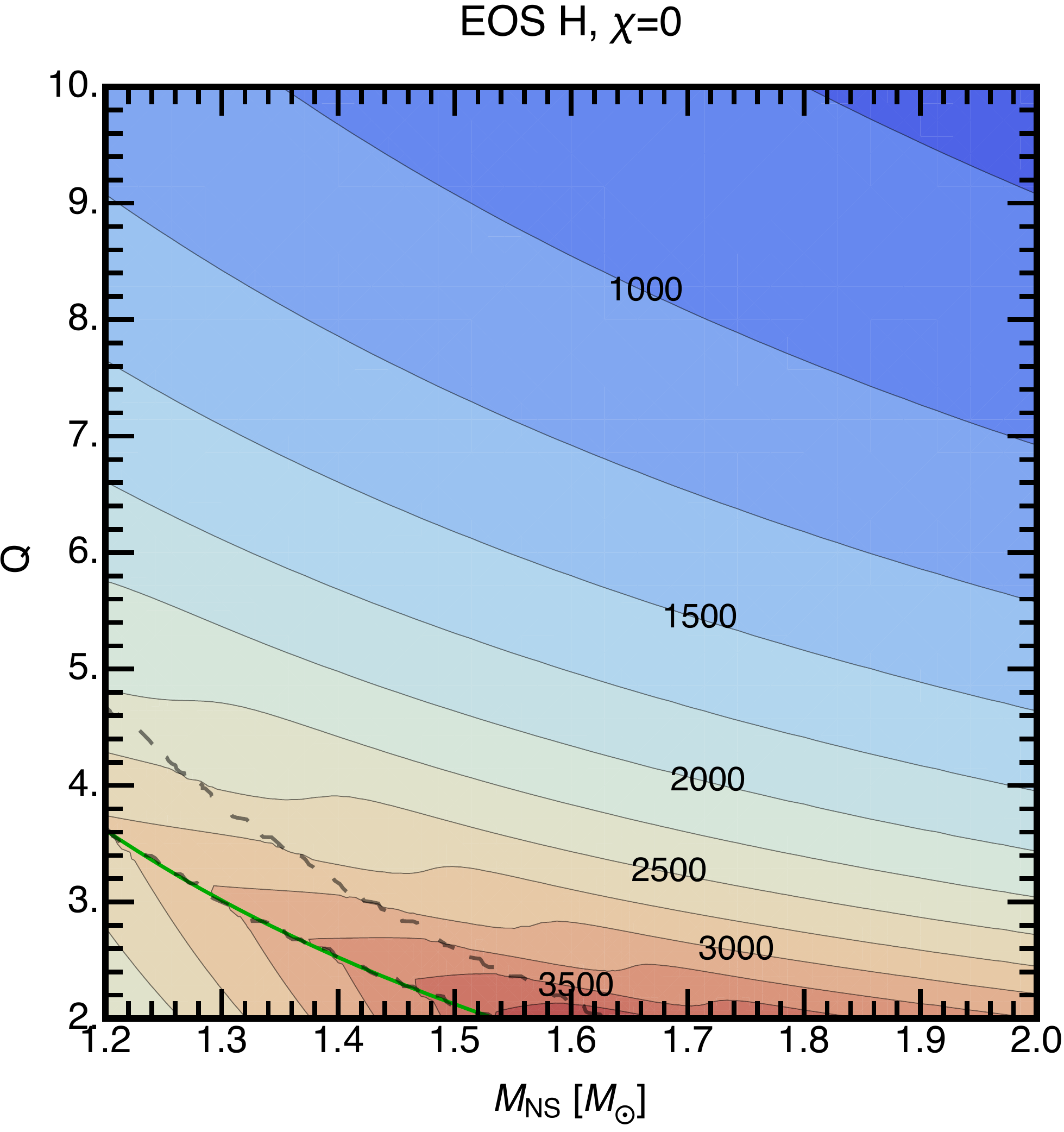}
    \\
    \includegraphics[width=\columnwidth,clip=true]{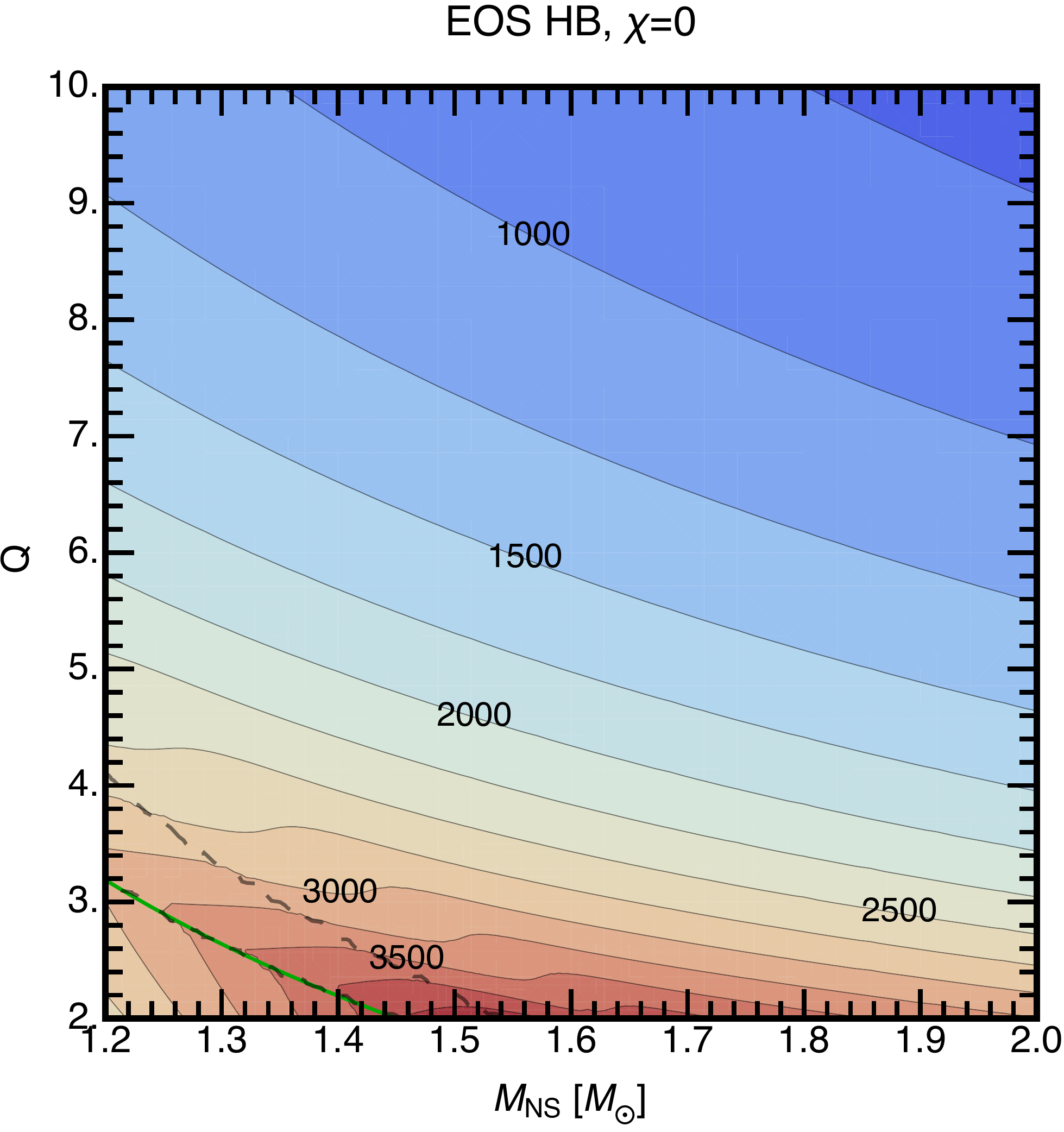}
    &
    \includegraphics[width=\columnwidth,clip=true]{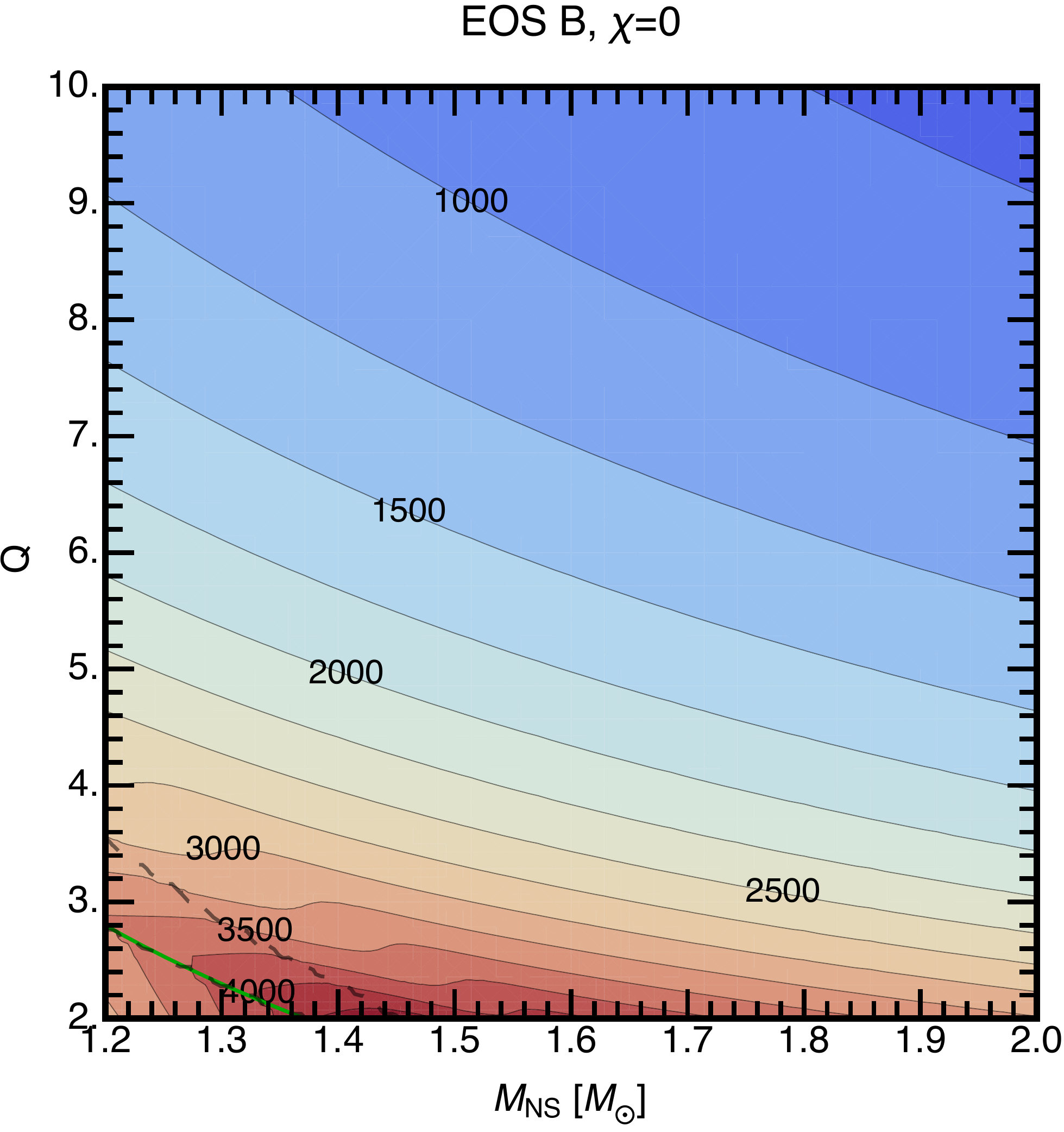}
  \end{tabular*}
  \caption{The cutoff frequency $\fCut$, as defined in
    Eq.\,(\ref{eq:fCut}), computed with our \nsbh \ac{GW} amplitude
    model.  We set the \ac{BH} spin parameter $\chi$ to zero and
    consider the \acp{EOS} 2H, H, HB, B.  The contour lines report
    $\fCut$ in Hz and have a spacing of $250\,$Hz.  The thick, green,
    continuous line is the location where $\mTorus$ vanishes, that is,
    where the left hand side of Eq.\,(\ref{eq:diskMass}) is zero.  The
    dashed lines in each panel divide the plane in three regions,
    explicitly labelled only in the top-right panel to avoid
    overcrowding the plots.  These are: a top-right region in which
    the \nsbh coalescences are nondisruptive, i.e.~$\fTide\geq\fRD$
    and $\mTorus=0$; a bottom-left one in which they are disruptive,
    i.e.~$\fTide<\fRD$ and $\mTorus>0$; and a middle region in which
    mildly disruptive coalescences occur, i.e.~$\fTide<\fRD$ and
    $\mTorus=0$, or $\fTide\geq\fRD$ and $\mTorus>0$.  This Figure
    should be compared to Fig.9 of Paper I: notice how the contour
    lines and the transitions between different regimes are smoother
    with the new model we formulate in this
    paper.\label{fig:fCutContours_chi0}}
\end{figure*}

\section{Applications}\label{sec:Applications}

\begin{figure*}[!tb]
  \begin{tabular*}{\textwidth}{c@{\extracolsep{\fill}}c@{\extracolsep{\fill}}c}
    \includegraphics[width=0.32\textwidth,clip=true]{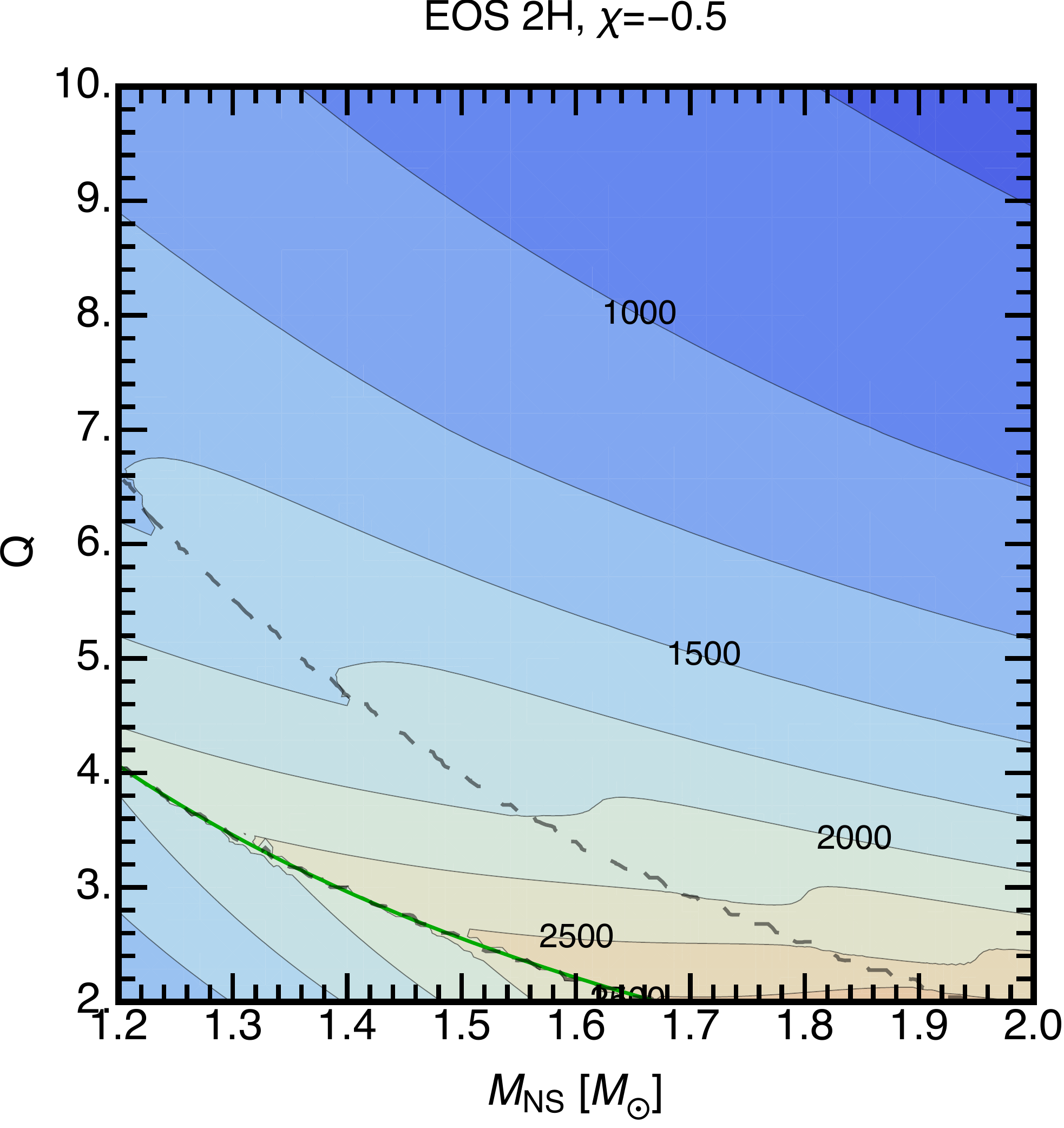}
    &
    \includegraphics[width=0.32\textwidth,clip=true]{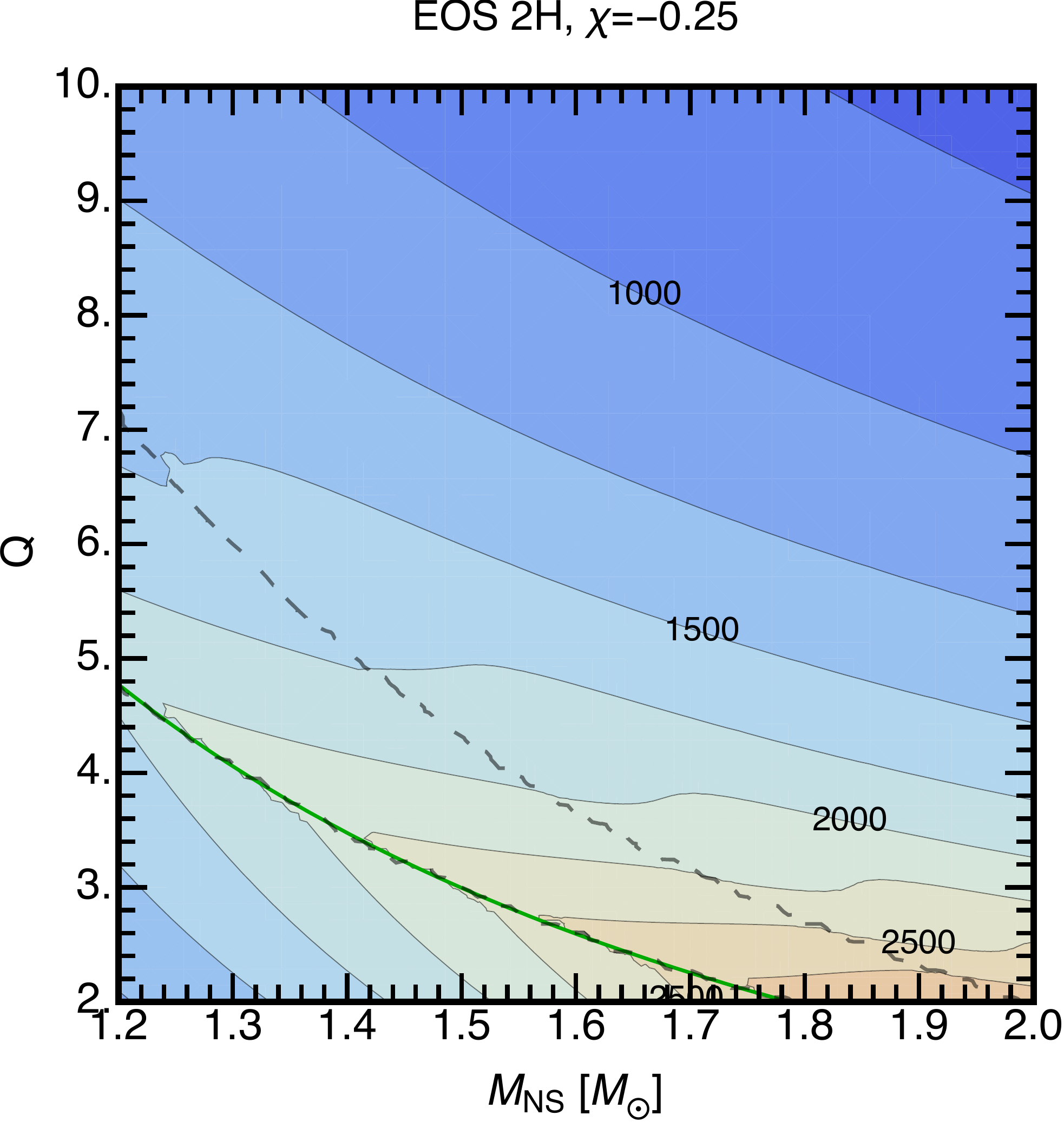}
    &
    \includegraphics[width=0.32\textwidth,clip=true]{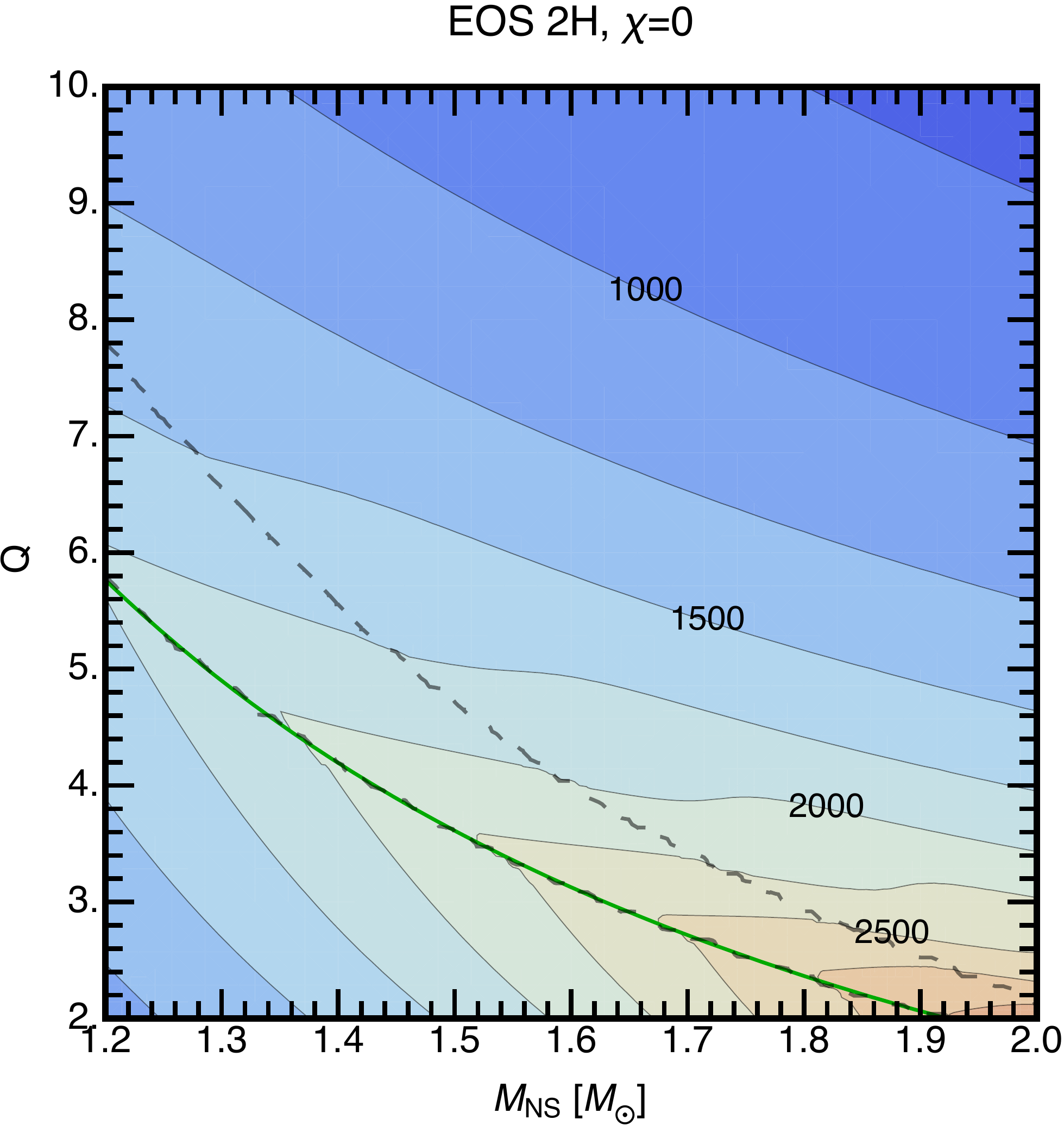}
    \\
    \includegraphics[width=0.32\textwidth,clip=true]{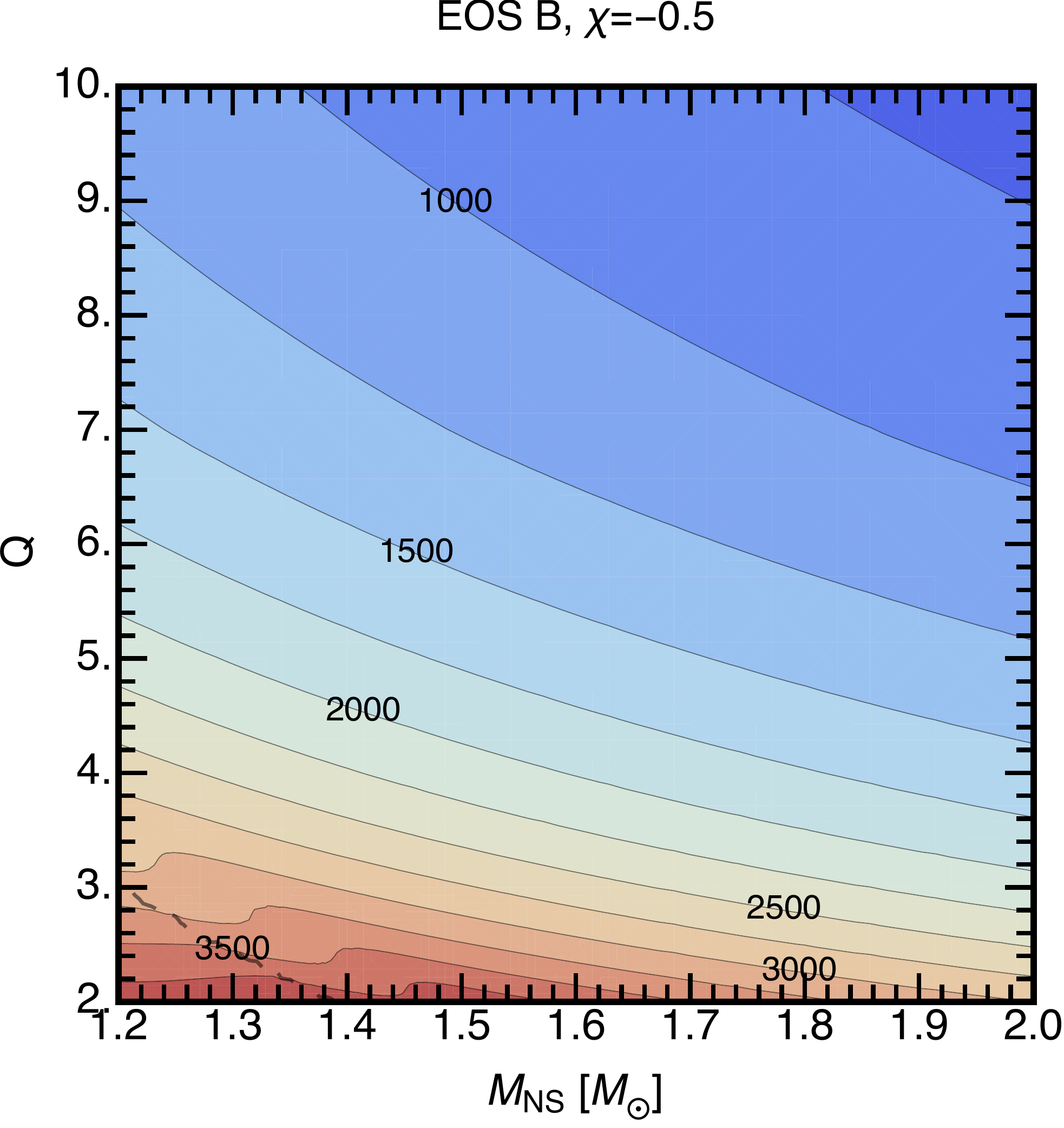}
    &
    \includegraphics[width=0.32\textwidth,clip=true]{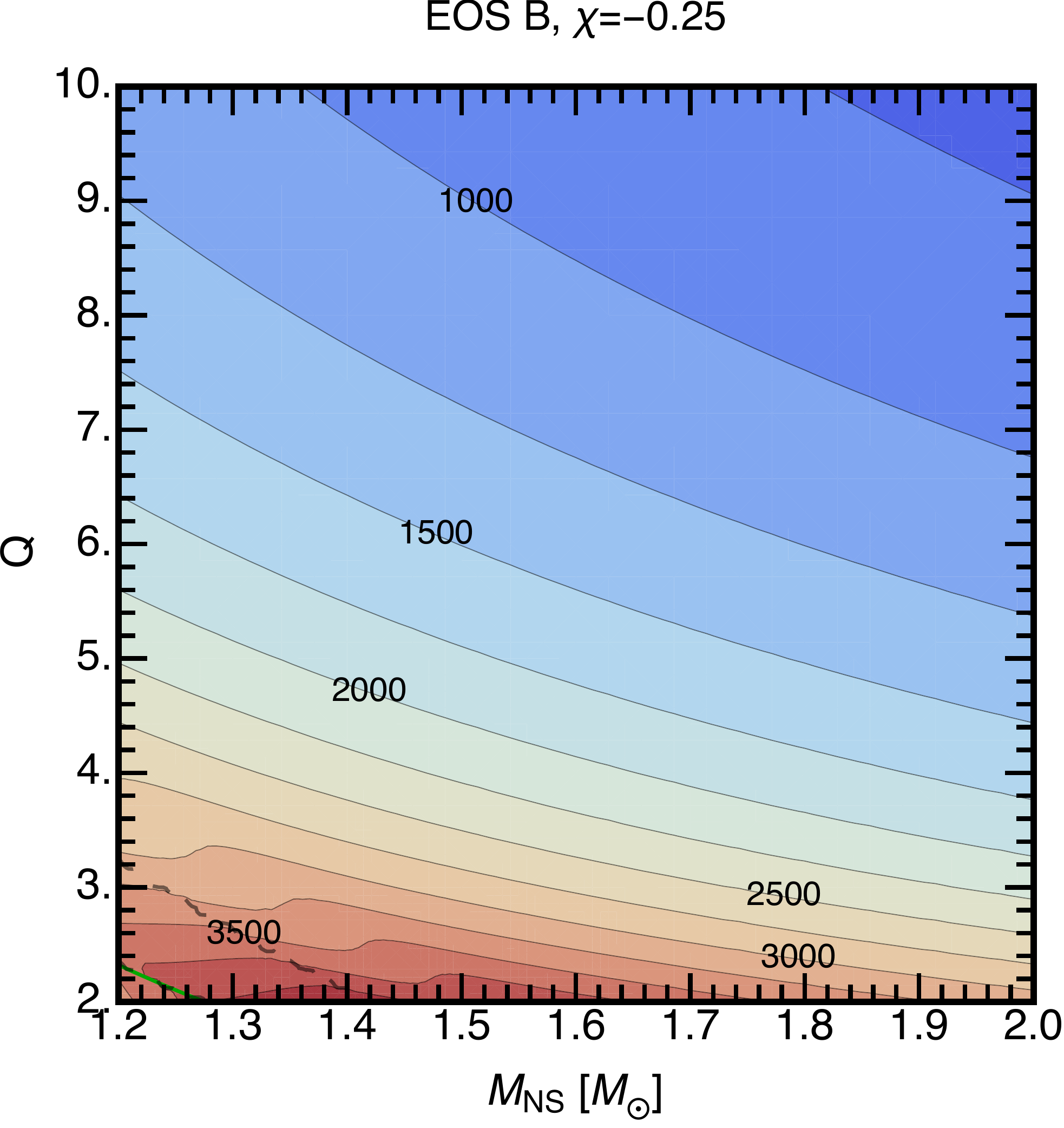}
    &
    \includegraphics[width=0.32\textwidth,clip=true]{FIG5d}
    \\
    \includegraphics[width=0.32\textwidth,clip=true]{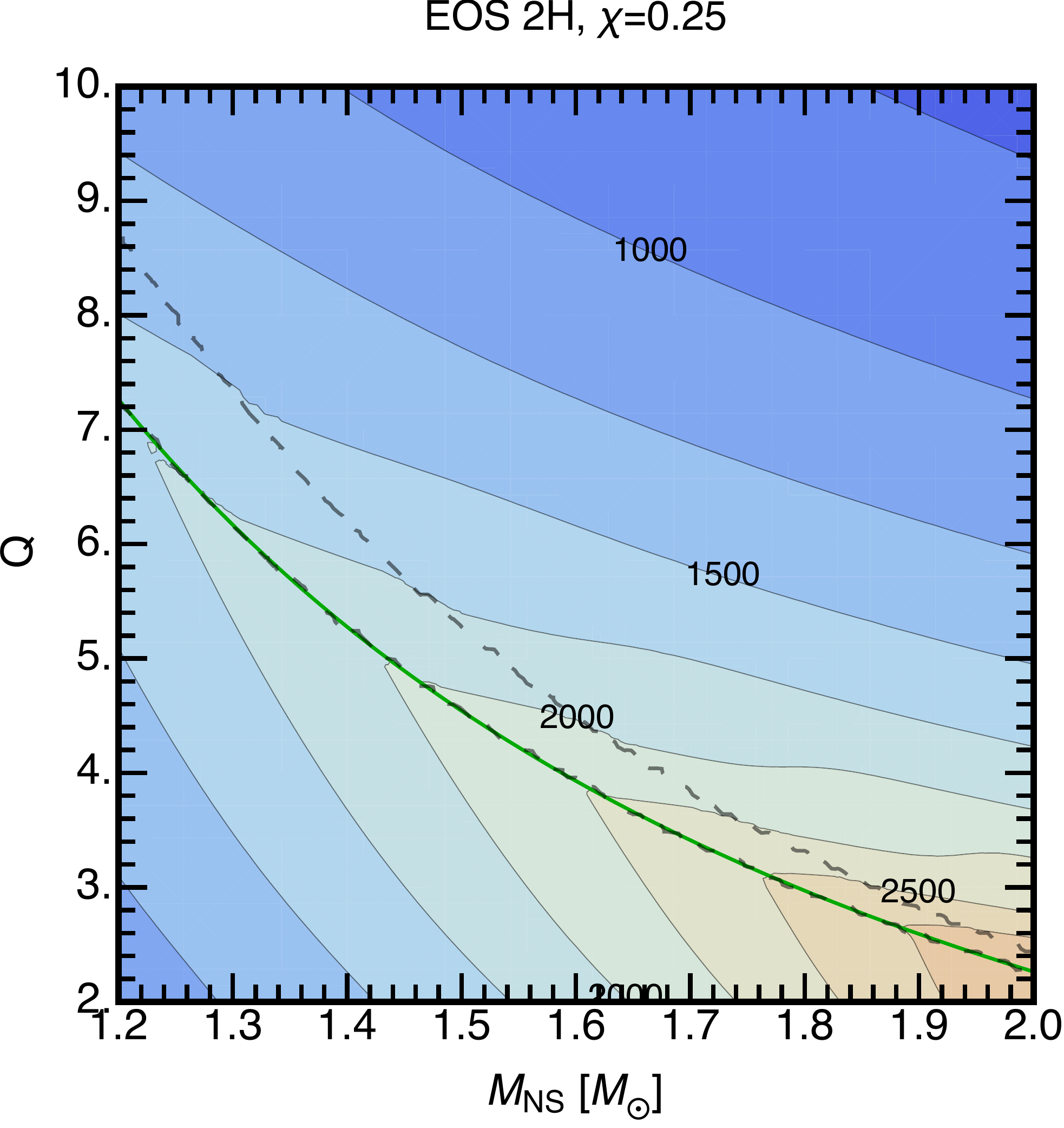}
    &
    \includegraphics[width=0.32\textwidth,clip=true]{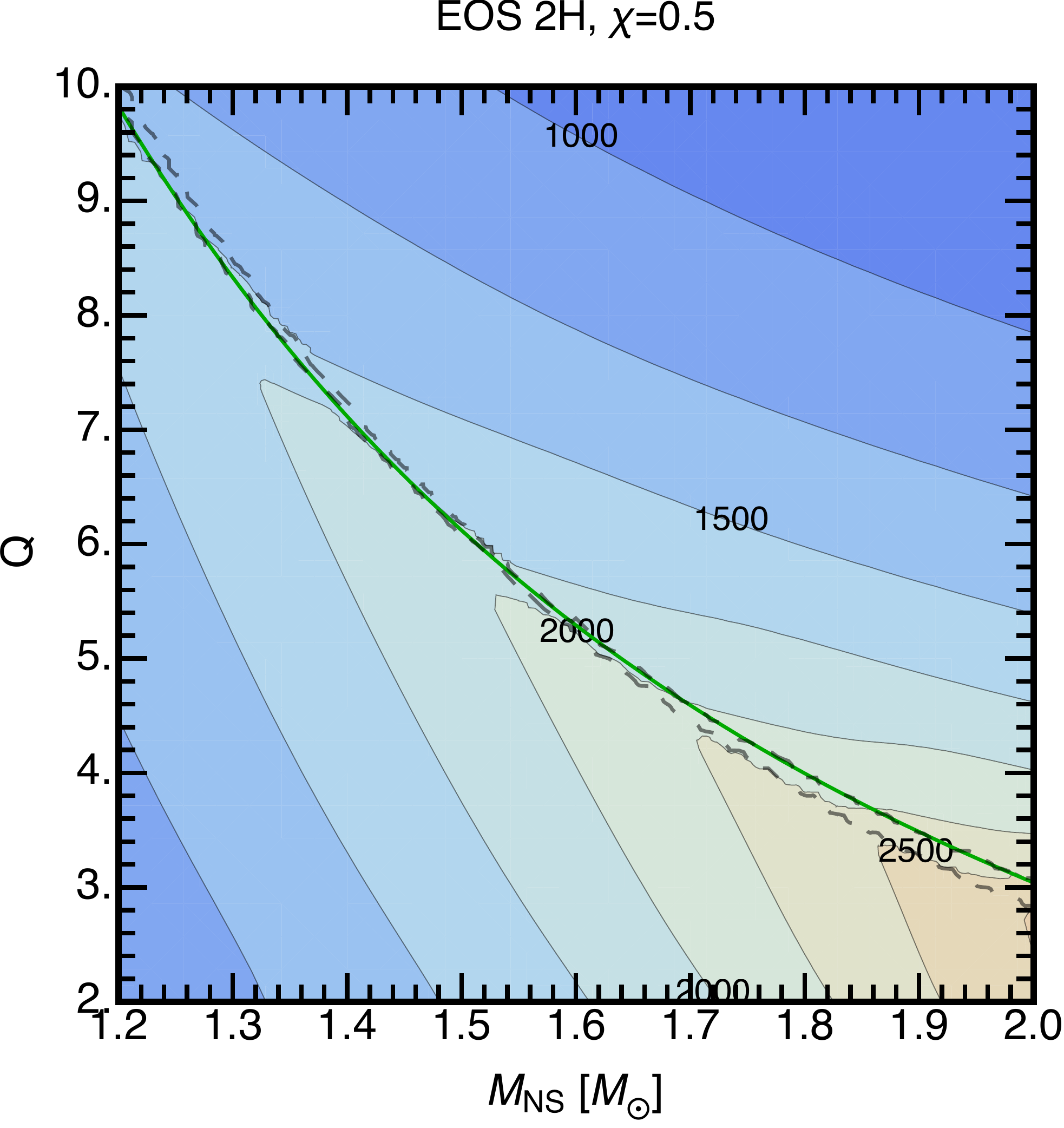}
    &
    \includegraphics[width=0.32\textwidth,clip=true]{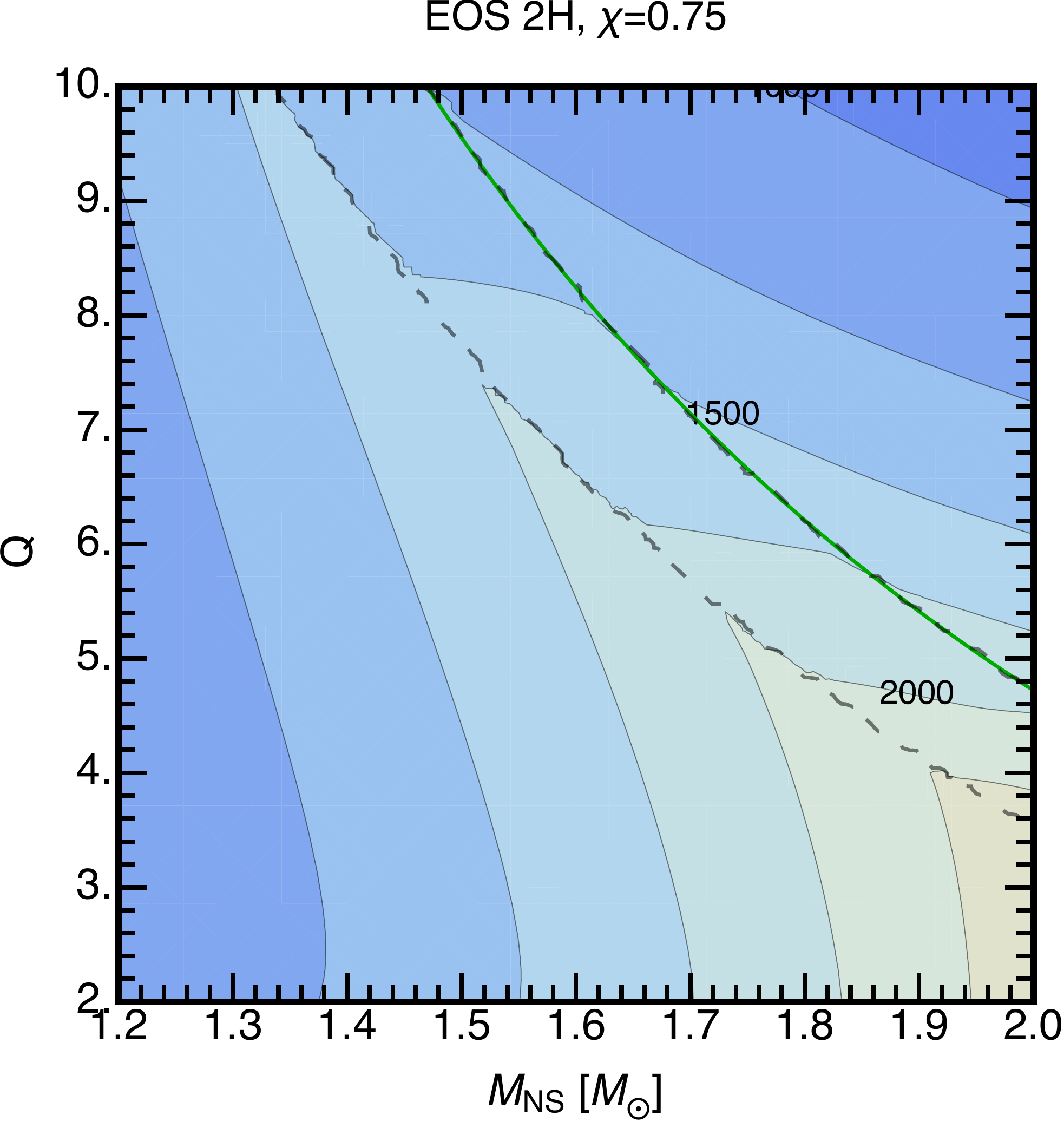}
    \\
    \includegraphics[width=0.32\textwidth,clip=true]{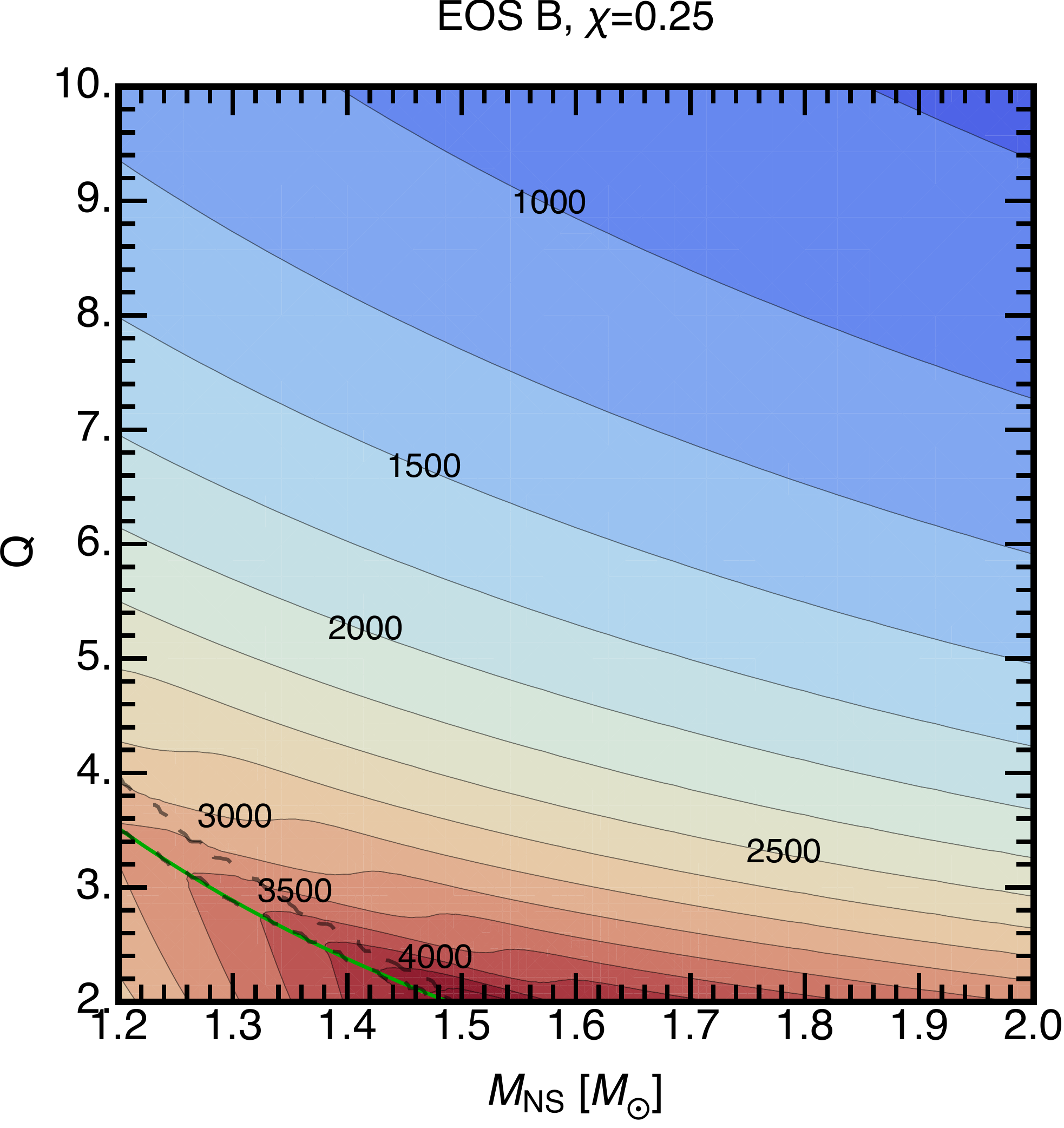}
    &
    \includegraphics[width=0.32\textwidth,clip=true]{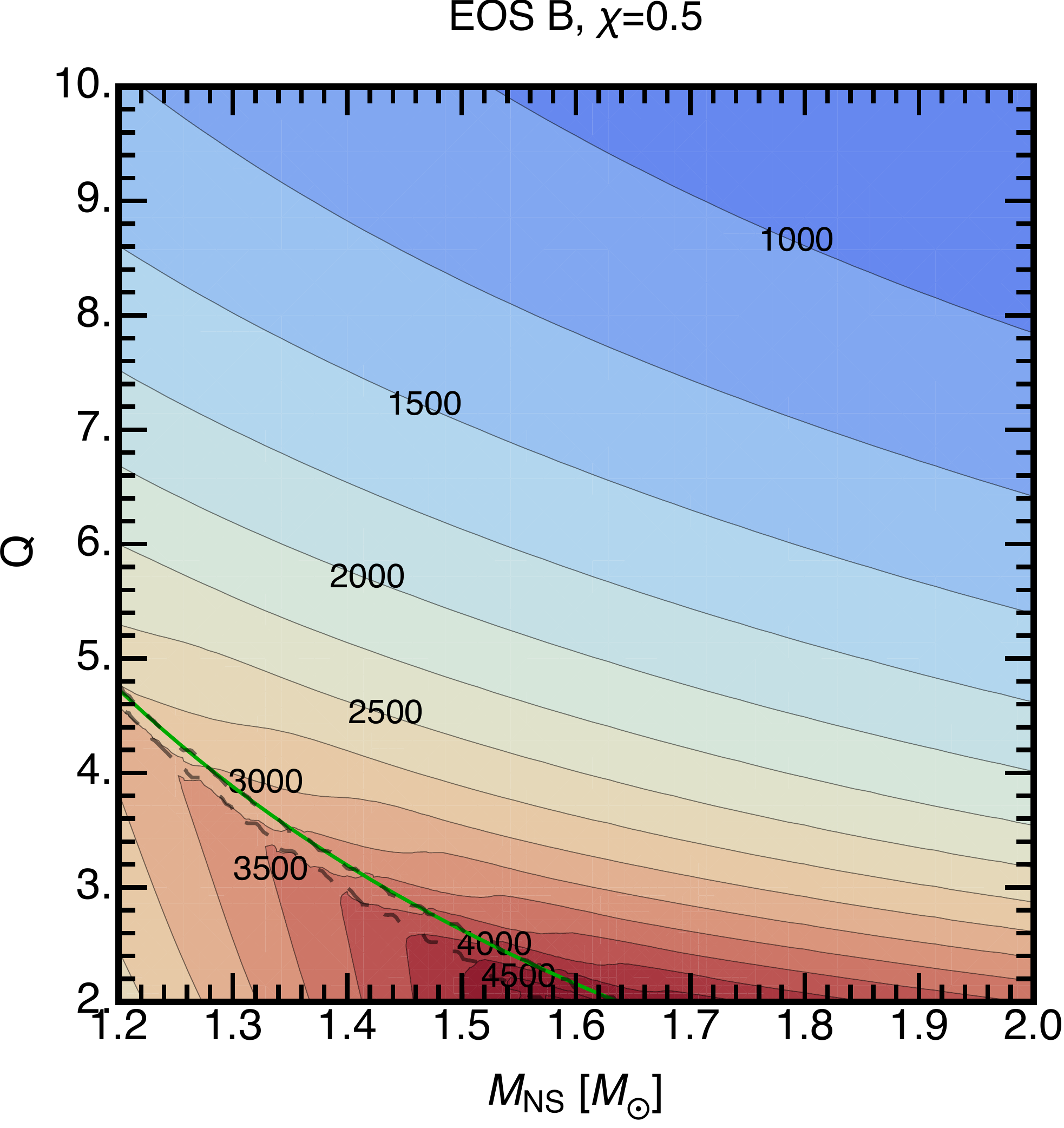}
    &
    \includegraphics[width=0.32\textwidth,clip=true]{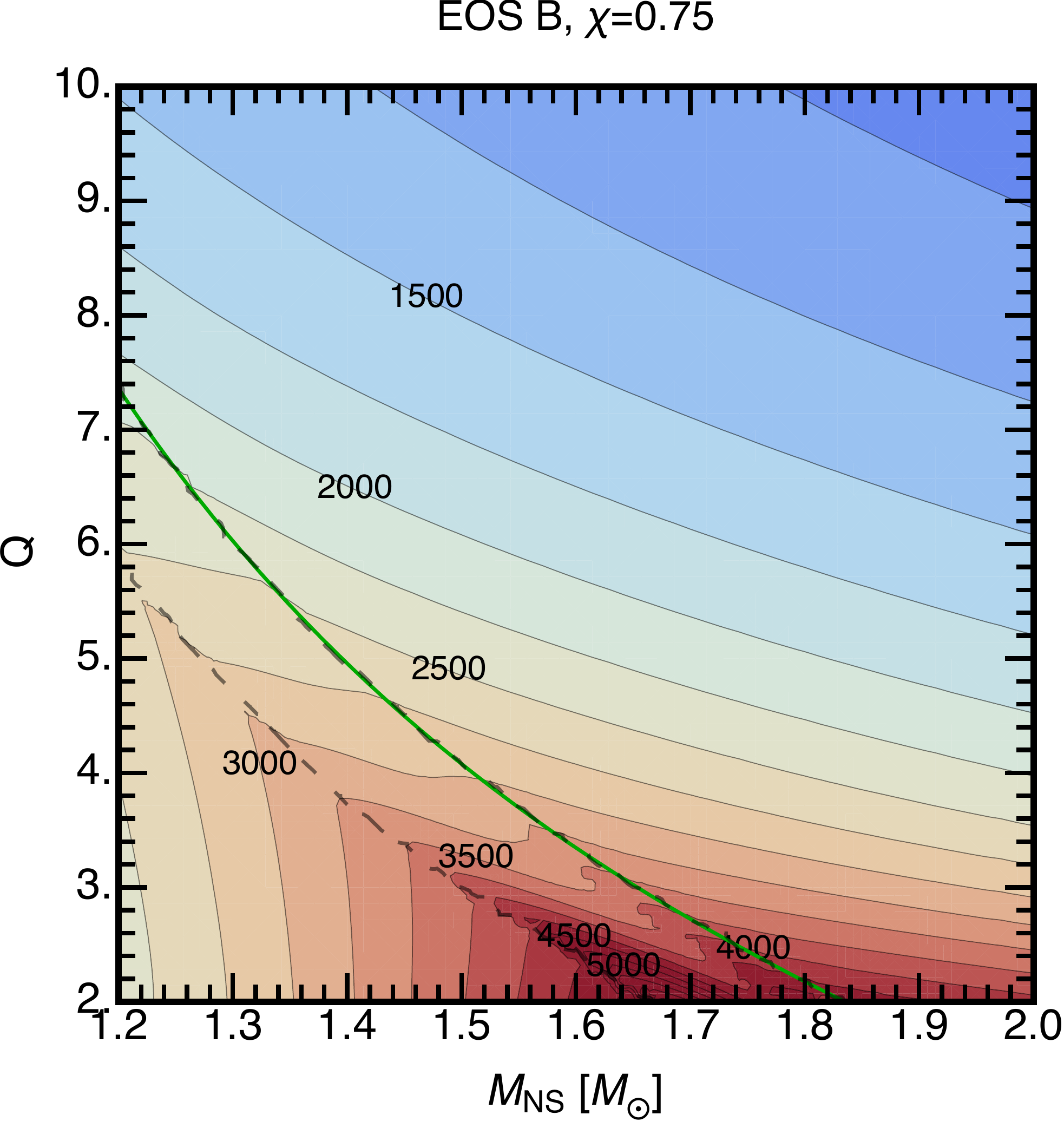}
    \\
  \end{tabular*}
  \caption{Same as Fig.~\ref{fig:fCutContours_chi0}, but focusing only
    on the ``extreme'' \acp{EOS} 2H and B. The \ac{BH} spins range
    between $-0.5$ and $0.75$ in steps of $0.25$, as indicated in each
    panel. \label{fig:fCutContours_spins}}
\end{figure*}

\subsection{The Cutoff Frequency}\label{sec:fCutContours}
The panels in Figure \ref{fig:fCutContours_chi0} show the cutoff
frequency $\fCut$ of nonspinning \nsbh binaries as a function of the
\ac{NS} mass $\mNS$ and the binary mass ratio $Q$ for the four
different \acp{EOS} (2H, H, HB and B) used to compute the hybrid
waveforms on which we built our phenomenological \ac{GW} amplitude
model.  This figure is an updated version of Figure 9 in Paper I.  The
contour line values are reported in Hz, with a $250\,$Hz spacing.  The
two dashed lines in each panel separate disruptive mergers
(bottom-left region), non-disruptive mergers (top-right region), and
mildly disruptive mergers (middle region in between the two lines).
We remind the reader that disruptive mergers correspond to the
conditions $\fTide<\fRD$ and $\mTorus>0$; nondisruptive mergers are
such that $\fTide>\fRD$ and $\mTorus=0$; and mildly disruptive mergers
do not fall into either of the previous categories.  To help the
reader, the three regions are explicitly indicated in the top-left
panel of the figure.  In all panels, a green line marks the boundary
between binaries with $\mTorus>0$ and those with $\mTorus=0$.  Notice
how these green lines overlap with the lower dashed lines: this
indicates that the mildly disruptive mergers in these panels are such
that the \ac{NS} is tidally disrupted, but no remnant torus is
formed. The main difference between these plots and those in Paper I
is that the behavior of $\fCut$ across the dashed transition lines is
now smoother, and that the contours are continuous.  This improvement
over the original model is due to the larger data set used here, which
allowed us to better tune our model in the mildly disruptive region.

The spin dependence of $\fCut$ is illustrated in Figure
\ref{fig:fCutContours_spins}, where the different panels refer to
initial \ac{BH} spin parameters $\chi\in \{-0.5,-0.25,0,0.25, 0.5,
0.75\}$ and we show two extreme cases for the \ac{EOS} (2H and B).  By
comparing the \ac{EOS} 2H and the \ac{EOS} B panels, we therefore get
an idea of the span of possible cutoff frequencies $\fCut$ at a given
\ac{BH} spin value.  As expected, the main differences occur in the
disruptive merger regions, as this is where the \ac{NS} \ac{EOS}
impacts the dynamical evolution of the binary.  The relative size of
the disruptive region grows with $\chi$, because larger spins increase
the likelihood of tidal disruption.  It is interesting to notice that
above $\chi\simeq 0.5$ the green torus mass boundary lines no longer
track the lower dashed lines, but the upper ones (cf. the panels with
$\chi=0.75$, and the \ac{EOS} B panel with $\chi=0.5$).  This means
that a small remnant torus is likely to be formed for mildly
disruptive mergers in these cases.  It also demonstrates that there is
indeed a need to split the phenomenological \ac{GW} model into four
subcases, as we do in this paper, at least until better analytical
predictions are available for $\mTorus$, $\fTide$, and $\fRD$, if one
wants to keep using these as tools to build the waveform model.

\begin{table*}[!tb]
  \caption{\label{tab:fit_coeffs} Values of the coefficients of the fits discussed in Eqs.\,(\ref{eq:DMDfitMNS}) and (\ref{eq:fCutFitMNS}).  The number below each coefficient symbol must be multiplied by the power of ten in square brackets on the right-hand side of the coefficient symbol.  The $g_{ijk}$'s are reported in $G=c=m_0=1$ units.}
  \resizebox{\textwidth}{!}{%
    \begin{tabular}{c@{\hspace{0.3cm}}c@{\hspace{0.3cm}}c@{\hspace{0.3cm}}c@{\hspace{0.3cm}}c@{\hspace{0.3cm}}c@{\hspace{0.3cm}}c@{\hspace{0.3cm}}c@{\hspace{0.3cm}}c@{\hspace{0.3cm}}c@{\hspace{0.3cm}}}
      \toprule[1.pt]
      \toprule[1.pt]
      \addlinespace[0.3em]
      $b_{00}$ $[10^{1}]$ & $b_{10}$ $[10^{1}]$ & $b_{01}$ $[10^{1}]$ & $b_{20}$ $[10^{1}]$ & $b_{11}$ $[10^{1}]$ & $b_{02}$ $[10^{1}]$ & $b_{30}$ $[10^{0}]$ & $b_{21}$ $[10^{0}]$ & $b_{12}$ $[10^{0}]$ & $b_{03}$ $[10^{-1}]$ \\
      $4.14730$ & $-5.70783$ & $2.57882$ & $2.91134$ & $-2.44263$ & $1.04225$ & $-5.26102$ & $6.28215$ & $-5.13944$ & $3.99706$ \\
      \addlinespace[0.2em]
      \midrule[1.pt]
      \addlinespace[0.4em]
      $g_{000}$ $[10^{-2}]$ & $g_{100}$ $[10^{-1}]$ & $g_{010}$ $[10^{-2}]$ & $g_{001}$ $[10^{-2}]$ & $g_{200}$ $[10^{-2}]$ & $g_{020}$ $[10^{-3}]$ & $g_{002}$ $[10^{-2}]$ & $g_{110}$ $[10^{-2}]$ & $g_{101}$ $[10^{-3}]$ & $g_{011}$ $[10^{-4}]$ \\
      $9.17677$ & $-1.39031$ & $-2.61399$ & $2.43286$ & $5.46375$ & $1.74490$ & $4.16418$ & $2.31878$ & $-7.49673$ & $5.65265$ \\
      \colrule
      $g_{300}$ $[10^{-3}]$ & $g_{030}$ $[10^{-5}]$ & $g_{003}$ $[10^{-3}]$ & $g_{210}$ $[10^{-3}]$ & $g_{120}$ $[10^{-4}]$ & $g_{201}$ $[10^{-4}]$ & $g_{102}$ $[10^{-2}]$ & $g_{021}$ $[10^{-4}]$ & $g_{012}$ $[10^{-3}]$ & $g_{111}$ $[10^{-3}]$ \\
      $-1.28721$ & $-6.88240$ & $-2.32757$ & $3.64301$ & $-4.05234$ & $-8.98986$ & $-1.87475$ & $5.50808$ & $-5.78858$ & $-8.55090$ \\
      \bottomrule[1.pt]
      \bottomrule[1.pt]
    \end{tabular}
  }
\end{table*}

\begin{figure*}[!tb]
  \begin{tabular*}{\textwidth}{c@{\extracolsep{\fill}}c}
    \includegraphics[width=\columnwidth,clip=true]{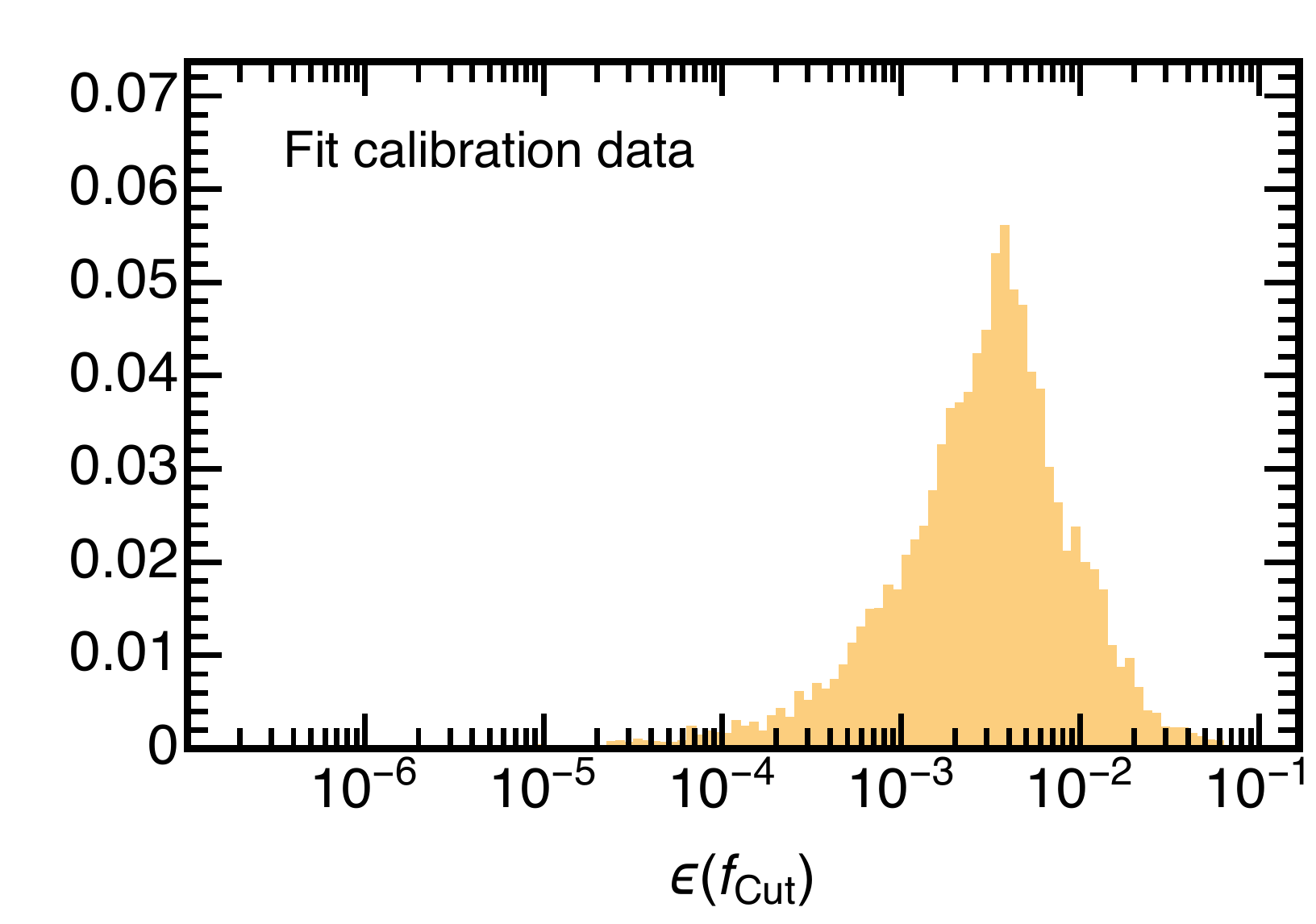}&
    \includegraphics[width=\columnwidth,clip=true]{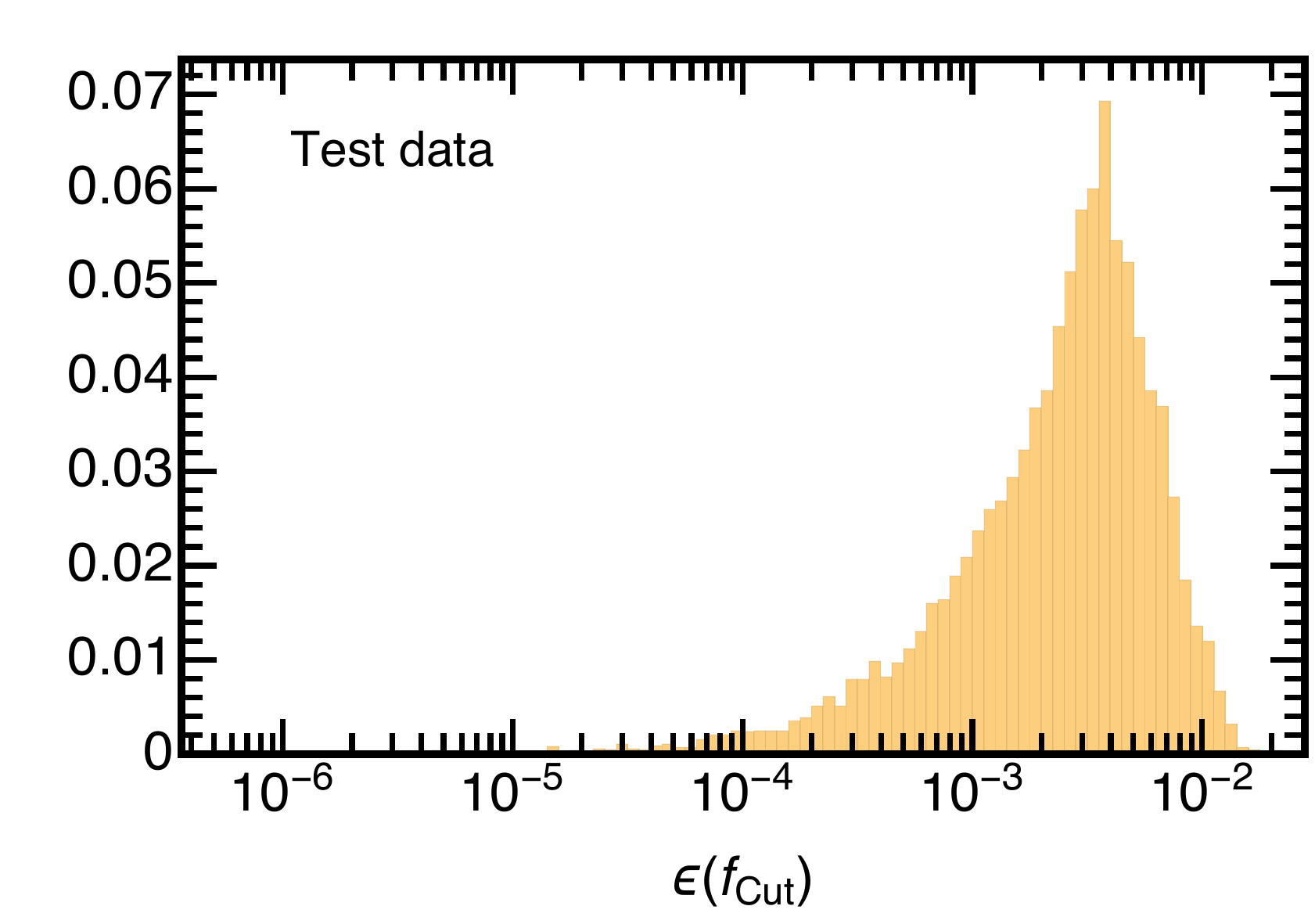}
  \end{tabular*}
  \caption{Distribution of relative error on $\fCut$ obtained by
    comparing the values given by the fit in Eq.\,(\ref{eq:fCutFit})
    to the $10^4$ \ac{EOS} 2H datapoints, used to produce the fit
    (left) and to $10^4$ \ac{EOS} B datapoints not involved in
    producing the fit (right). \label{fig:fCutFitErrors}}
\end{figure*}

\subsection{Phenomenological Fits}\label{sec:fCutFit}
As shown in a companion paper, the results presented above can be used
to construct simple phenomenological formulas to determine (1) whether
an \nsbh binary is disruptive, mildly disruptive, or nondisruptive,
and (2) the \ac{GW} cutoff frequency for disruptive mergers, due to
the \ac{NS} tidal disruption. Details are given in~\cite{PRLprep}, and
here we only review the main results.

\subsubsection{Disruption Criterion}
The contours that separate \nsbh binaries with a disruptive fate from
those with a mildly disruptive or nondisruptive fate in
Figs.~\ref{fig:fCutContours_chi0} and \ref{fig:fCutContours_spins} may
be fitted in several ways as a function of the physical parameters of
the binary. The critical binary mass ratio $Q_{\rm D}=Q_{\rm
  D}(\mathcal{C},\chi)$ below which mergers are disruptive is an
approximately ``universal'' (i.e.~\ac{EOS}-independent) function that
is well fitted by:
\begin{align}%
  \label{eq:DMDfit}
  Q_{\rm D} &= \sum_{\substack{i,j=0\\i+j\leq 3}}^3 a_{ij}
  \mathcal{C}^i\chi^j\,,
\end{align}%
where the $a_{ij}$'s are constants (see~\cite{PRLprep} for details).

In building template banks for \ac{GW} detection and for other
applications, it may be useful to know $Q_{\rm D}$ as a function of
the \ac{NS} mass for \acp{NS} with large radii, as this is the case in
which the \ac{GW} emission from \nsbh systems differs the most from
\ac{BH}-\ac{BH} binaries.  A fit of the disruptive boundary for the 2H
\ac{EOS} yields
\begin{align}%
  \label{eq:DMDfitMNS}
  Q_{\rm D} &= \sum_{\substack{i,j=0\\i+j\leq 3}}^3 b_{ij}
  \bmNS^i\chi^j\,,
\end{align}%
where $\bmNS=\mNS/\mSun$ and the coefficients take the values reported
in Table \ref{tab:fit_coeffs} in units of $G=c=m_0=1$.  We remark
again, that this fit gives a lower limit on $Q_{\rm D}$, as it was
obtained using an exceptionally stiff \ac{EOS}.

\subsubsection{Cutoff Frequency Fitting Formula}
When tidal disruption occurs, as determined via
Eq.\,(\ref{eq:DMDfit}), our phenomenological model allows us to
determine a formula that provides the \ac{GW} cutoff frequency
analytically, as follows.  We consider the 2H \ac{EOS}, generate a set
of $10^4$ random disruptive mergers, compute $\fCut$ for each \nsbh
binary according to the definition in Eq.\,(\ref{eq:fCut}), and
finally fit the data thus obtained.  In order to select disruptive
mergers, we randomly sample the parameter space in the ranges
$\mNS/M_\odot\in[1.2,2.83]$ and $Q\in[2,10]$, $\chi\in[-0.5,0.75]$; we
verify whether the sampled point corresponds to a disruptive binary,
as defined just above Eq.\,(\ref{eq:PhenoMixedAmpD}), and keep the
point if it does. The whole process is repeated until we have the
desired set of $10^4$ disruptive binaries.  While the maximum \ac{NS}
mass for the 2H \ac{EOS} is $\sim 2.83M_\odot$, the maximum \ac{NS}
mass in our sample of disruptive \nsbh mergers is $\sim
2.28M_\odot$.
The resulting mass interval $\mNS/M_\odot\in[1.2,2.28]$ corresponds to
the compactness interval $0.117 \leq \mathcal{C}\leq 0.221$.  With
this set of disruptive cutoff frequency data in hand, we fit $\fCut$
in terms of the \nsbh binary parameters using the ansatz
\begin{align}%
  \label{eq:fCutFit}
  \fCut &= \sum_{\substack{i,j=0\\i+j+k\leq 3}}^3 f_{ijk}
  \mathcal{C}^i Q^j \chi^k\,,
\end{align}%
where the fitting coefficients can be found in~\cite{PRLprep}.  The
relative error distribution for this fit with respect to the original
data points is shown in the left panel of Figure
\ref{fig:fCutFitErrors}.  Notice that the peak of the distribution is
below the percent level: the relative error for $68$\%, $95$\%, and
$99.7$\% of the points is $0.47$\%, $1.5$\%, and $4.9$\%,
respectively.

As a consistency check for this fitting formula, we draw a separate
sample of $10^4$ disruptive mergers, compute the \ac{GW} amplitude
cutoff frequency for each binary, and determine the relative errors of
the fit just discussed against these ``test'' binaries.  This time we
use \ac{EOS} B to construct our ``test'' sample and we lower the
maximum allowed \ac{NS} mass to $2M_\odot$, as this is approximately
the maximum $\mNS$ for this \ac{EOS}.  The compactness now ranges
between $\sim 0.161$ and $\sim 0.225$.  The result of this test is
reported in the right panel of Figure \ref{fig:fCutFitErrors}, where
we show the relative error distribution for the fit in
Eq.\,(\ref{eq:fCutFit}) with respect to the ``test'' set of binary
mergers populated using \ac{EOS} B.  Remarkably, the maximum relative
error is $2.2\%$, $97.6$\% of the points have a relative error that is
smaller than $1$\%, and the peak of the distribution is once again
below the percent level. The relative errors for the \ac{EOS} B
``test'' set are even better than for the ``calibration'' set of
\ac{EOS} 2H, because \ac{EOS} B covers a narrower range in compactness
relative to \ac{EOS} 2H.  Furthermore, the fit of
Eq.\,\eqref{eq:fCutFit} is effectively EOS-independent (or
``universal''), at least within the parameter space region in which
our model was calibrated.

For \ac{GW} data analysis purposes, we also performed a fit of the 2H
\ac{EOS} $\fCut$ data in terms of the \ac{NS} mass, rather than its
compactness:
\begin{align}%
  \label{eq:fCutFitMNS}
  \fCut &= \sum_{\substack{i,j=0\\i+j+k\leq 3}}^3 g_{ijk} \bmNS^i Q^j \chi^k\,,
\end{align}%
where the coefficients $g_{ijk}$ are listed in Table
\ref{tab:fit_coeffs}.  The resulting error distribution is similar to
the one shown in the left panel of Figure \ref{fig:fCutFitErrors}, but
now the relative errors with respect to the \ac{EOS} B ``test'' data
are much higher than those in the right panel of
Fig.~\ref{fig:fCutFitErrors} and range from $54$\% to $65$\%.  In
other words, the mass fit in Eq.\,(\ref{eq:fCutFitMNS}) is \emph{not}
EOS-universal. The fit is still useful, as it provides a lower limit
to the \ac{GW} amplitude cutoff frequency$\fCut$ as a function of
$\mNS$, $Q$ and $\chi$.

\section{Conclusions}\label{sec:conclusions}
In this paper we have extended the phenomenological gravitational
waveform amplitude model of Paper I to \nsbh binaries in which the
\ac{BH} has spin either aligned or antialigned with the orbital
angular momentum. We have classified binaries in four broad
categories: ``disruptive,'' ``nondisruptive,'' and ``mildly
disruptive'' with and without a torus remnant.

The phenomenological tools developed in this paper can be used in
various contexts, as detailed elsewhere~\cite{PRLprep}. Our
predictions for the \ac{GW} amplitude and for the cutoff frequency may
be used to improve the template banks currently used in \nsbh
searches, and they could also be exploited to build new \nsbh
phenomenological \ac{GW} phase models and \ac{EOB} models.  More
accurate gravitational waveforms improve our chances of detecting
\ac{GW} signals and of extracting information from them.  For example
they could provide better constraints on the \ac{NS}
\ac{EOS}~\cite{Read:2009yp} and possibly even on the underlying theory
of gravity~\cite{Berti:2015itd}. Furthermore, our work allows us to
pin-point binaries in which tidal effects are relevant.  These cause
the \ac{GW} signal to deviate significantly from a \bbh one and,
possibly, lead to the emission of electromagnetic radiation.  The
latter may either be in the form of a relativistic jet launched by a
hot, massive ($\gtrsim 0.01\mSun$) disk produced by the tidal
disruption of the \ac{NS} (a scenario that could explain the duration,
energetics, and estimated event rates of
\acp{SGRB}~\cite{Paczynski:1991aq,Nakar:2007yr,Berger:2013jza}), or in
the form of isotropical radiation, i.e.~macronov\ae/kilonov\ae,
powered by decay heat of unstable $r$-process elements and by
nonthermal radiation from electrons accelerated at blast waves between
the merger ejecta and the interstellar medium~\cite{Kyutoku2013,
  Kyutoku2015, LiPaczynski1998, Kulkarni2005, Metzger2010, Nakar2011,
  Kisaka2015}.  For these reasons, our model can have an impact on
multimessenger searches targeting \ac{GW}, electromagnetic, and
neutrino radiation, as well as important applications in the
interpretation of future multimessenger observations.

All our predictions are clearly affected by systematics in the initial
data for the numerical simulations we used, in the numerical
evolutions, in the phenomenological model itself and the tools it
relies on~\cite{Foucart:2012nc, Pannarale2012}, and in the fitting
procedures. We expect these errors to increase when the model is
extrapolated beyond the parameter space in which it was tuned.  Future
work should extend and improve our model in order to include not only
the \ac{GW} frequency domain amplitude, but also its phase. It should
also consider larger values of the \ac{BH} spin, nonzero \ac{NS} spins
and (most importantly) precession effects that occur when the \ac{BH}
spin is not aligned with the orbital angular momentum (see
e.g.~\cite{Foucart:2012vn,Kawaguchi:2015bwa}).  Further, any
improvements in the underlying phenomenological \bbh model can and
should be included in our framework for \nsbh systems.  In particular,
the recent ``PhenomD'' \bbh model~\cite{PhenomDa, PhenomDb} is
calibrated to hybrid effective-one-body waveforms that use
numerical-relativity simulations with mass ratios up to $1$:$18$, and
\ac{BH} dimensionless spin parameters up to $\sim 0.85$ ($0.98$ in the
equal-mass case).  This model is an improvement with respect to the
``PhenomC'' model, which was calibrated up to mass ratios of $1$:$4$,
and resolves the technical limitations that may be encountered when
using it for \bbh systems with mass ratio higher than $1$:$20$ and
$|\chi|>0.9$.  We plan to address all of these issues in the near
future.

{\bf \em Acknowledgements.}
This work was supported by STFC grant No.~ST/L000342/1, by the
Japanese Grant-in-Aid for Scientific Research (21340051, 24244028),
and by the Grant-in-Aid for Scientific Research on Innovative Area
(20105004).  E.B.~is supported by NSF CAREER Grant PHY-1055103 and by
FCT contract IF/00797/2014/CP1214/CT0012 under the IF2014 Programme.
K.K.~is supported by the RIKEN iTHES project.  B.L.~was supported by
NSF grants PHY-1305682, PHY-1205835, and AST-1333142.  F.P.~wishes to
thank Alex Nielsen, Alessandra Buonanno, Stephen Fairhurst, Tanja
Hinderer, and Bangalore Sathyaprakash for interesting discussions
throughout the development of this work, along with Elena Pannarale
for all her support.


\bibliographystyle{apsrev4-1-noeprint}
\bibliography{PhenoMixedSpinGWs}

\end{document}